\documentclass[
  aip,
  amsmath, amssymb,
  reprint,
  author-year,
]{revtex4-1}

\usepackage[utf8]{inputenc}
\usepackage[T1]{fontenc}
\usepackage{graphicx}
\usepackage{dcolumn}
\usepackage{mathptmx}
\usepackage{etoolbox}
\usepackage{aas}
\usepackage{bm}

\makeatletter
\def\@email#1#2{%
 \endgroup
 \patchcmd{\titleblock@produce}
  {\frontmatter@RRAPformat}
  {\frontmatter@RRAPformat{\produce@RRAP{*#1\href{mailto:#2}{#2}}}\frontmatter@RRAPformat}
  {}{}
}%
\makeatother
\begin{document}

\preprint{AIP/123-QED}

\title[Slow wind belt in the quiet solar corona]{Slow wind belt in the quiet solar corona}

\author{E. Antonucci}
\email{ester.antonucci@inaf.it}
\affiliation{INAF - Astrophysical Observatory of Torino, Via Osservatorio, 20, 10025, Pino To.se, Torino, Italy}

\author{C. Downs}
\affiliation{Predictive Science Inc., 9990 Mesa Rim Rd., CA 92121, San Diego, CA, USA}

\author{G. E. Capuano}
\affiliation{INAF - Astrophysical Observatory of Catania, Via S. Sofia 78, 95123, Catania, Italy}

\author{D. Spadaro}
\affiliation{INAF - Astrophysical Observatory of Catania, Via S. Sofia 78, 95123, Catania, Italy}

\author{R. Susino}
\affiliation{INAF - Astrophysical Observatory of Torino, Via Osservatorio, 20, 10025, Pino To.se, Torino, Italy}

\author{D. Telloni}
\affiliation{INAF - Astrophysical Observatory of Torino, Via Osservatorio, 20, 10025, Pino To.se, Torino, Italy}

\author{V. Andretta}
\affiliation{INAF - Astronomical Observatory of Capodimonte, Salita Moiariello 16, 80131, Naples, Italy}

\author{V. Da Deppo}
\affiliation{IFN-CNR, Via Trasea, 7, 35131, Padova, Italy}

\author{Y. De Leo}
\affiliation{Max-Planck-Institut f\"ur Sonnensystemforschung, Justus-von-Liebig-Weg 3, 37077, G\"ottingen, Germany}
\affiliation{University of Catania, Via S. Sofia 78, 95123, Catania, Italy}

\author{S. Fineschi}
\affiliation{INAF - Astrophysical Observatory of Torino, Via Osservatorio, 20, 10025, Pino To.se, Torino, Italy}

\author{F. Frassetto}
\affiliation{IFN-CNR, Via Trasea, 7, 35131, Padova, Italy}

\author{F. Landini}
\affiliation{INAF - Astrophysical Observatory of Torino, Via Osservatorio, 20, 10025, Pino To.se, Torino, Italy}

\author{G. Naletto}
\affiliation{University of Padova -- Physics and Astronomy Department ``Galileo Galilei'', Via F. Marzolo 8, 35131, Padua, Italy}
\affiliation{IFN-CNR, Via Trasea, 7, 35131, Padua, Italy}

\author{G. Nicolini}
\affiliation{INAF - Astrophysical Observatory of Torino, Via Osservatorio, 20, 10025, Pino To.se, Torino, Italy}

\author{M. Pancrazzi}
\affiliation{INAF - Astrophysical Observatory of Torino, Via Osservatorio, 20, 10025, Pino To.se, Torino, Italy}

\author{M. Romoli}
\affiliation{University of Firenze, Via Sansone, 1, 50019, Sesto Fiorentino, Florence, Italy}

\author{M. Stangalini}
\affiliation{ASI - Agenzia Spaziale Italiana, Via del Politecnico s.n.c., 00133, Rome, Italy}

\author{L. Teriaca}
\affiliation{Max-Planck-Institut f\"ur Sonnensystemforschung, Justus-von-Liebig-Weg 3, 37077, G\"ottingen, Germany}

\author{M. Uslenghi}
\affiliation{INAF - Institute for Space Astrophysics and Cosmic Physics, Via Alfonso Corti 12, 20133, Milan, Italy}

\date{\today}

\begin{abstract}
    The slow solar wind belt in the quiet corona, observed with the Metis coronagraph on board Solar Orbiter on May 15, 2020, during the activity minimum of the cycle 24, in a field of view extending from 3.8~$R_\odot$ to 7.0~$R_\odot$, is formed by a slow and dense wind stream running along the coronal current sheet, accelerating in the radial direction and reaching at 6.8~$R_\odot$ a speed within 150~km~s$^{-1}$ and 190~km~s$^{-1}$, depending on the assumptions on the velocity distribution of the neutral hydrogen atoms in the coronal plasma.
    The slow stream is separated by thin regions of high velocity shear from faster streams, almost symmetric relative to the current sheet, with peak velocity within 175~km~s$^{-1}$ and 230~km~s$^{-1}$ at the same coronal level.
    The density-velocity structure of the slow wind zone is discussed in terms of the expansion factor of the open magnetic field lines that is known to be related to the speed of the quasi-steady solar wind, and in relation to the presence of a web of quasi separatrix layers, S-web, the potential sites of reconnection that release coronal plasma into the wind.
    The parameters characterizing the coronal magnetic field lines are derived from 3D MHD model calculations.
    The S-web is found to coincide with the latitudinal region where the slow wind is observed in the outer corona and is surrounded by thin layers of open field lines expanding in a non-monotonic way.
\end{abstract}

\maketitle

\section{\label{sec:intro}Introduction}

The UltraViolet Coronagraph Spectrometer \citep[UVCS;][]{kohl1995} on the Solar and Heliospheric Observatory \citep[SOHO;][]{domingo1995} was the first coronagraph to observe the solar wind propagating in the atmosphere of the Sun and characterize the wind physical properties in its early development.
This significant advance was made possible thanks to the adoption of a novel spectroscopic technique based on the Doppler dimming of the ultraviolet light resonantly scattered by coronal ions and atoms \citep{withbroe1982,noci1987}.
The main UVCS results on the solar wind in the corona are reported in several reviews over the past decades \citep[e.g.,][]{abbo2016,cranmer2017,antonucci2020a}.

UVCS was designed to prioritize observations at high spatial and spectral resolutions of the corona in the ultraviolet regions that include spectral lines, formed by resonant scattering of the photons emitted by the chromosphere and transition region, and affected by Doppler dimming in an expanding corona.
In order to obtain the excellent spectroscopic capabilities of UVCS, data were collected in a narrow instantaneous field of view at different times and at different positions.
Hence UVCS maps of the wind outflow velocity in corona are reconstructed under the hypothesis of quasi-steady conditions from data necessarily lacking of spatial continuity and temporal simultaneity.

UVCS allowed the discovery of fundamental properties of both proton and oxygen components of the solar wind.
For instance, by analyzing in the coronal hole regions the profiles of the oxygen spectral lines -- which allow to extend the Doppler dimming diagnostics to higher outflow velocities than those accessible by studying the hydrogen H~{\sc i} Ly$\alpha$ line -- it was possible to obtain compelling evidence for the presence of energy deposition across the coronal magnetic field \citep{kohl1998,cranmer1999a,cranmer1999b} and to trace the oxygen component of the fast solar wind out to 5~$R_\odot$ where a speed $\gtrsim 550$~km~s$^{-1}$ is reached \citep{telloni2007a}.
In the case of the proton component, the outflow velocity in the polar holes, discussed by \citet{cranmer2020} on the basis of UVCS observations and solar wind models, is within 300-400~km~s$^{-1}$ close to 4~$R_\odot$.
The analysis of the H~{\sc i} Ly$\alpha$ radiation detected during solar minimum in June 1997 shows that the boundary between the fast wind originating in the polar regions and the slower wind zone is approximately at $\pm 30^\circ$ latitude from the solar equator \citep{dolei2018}.

The Metis coronagraph \citep{antonucci2020b} on the Solar Orbiter mission \citep{muller2020}, is based on the UVCS heritage and it has been primarily designed to complement UVCS by prioritizing the study of the dynamics and evolution the solar wind with unprecedented temporal and spatial resolution over annular regions of the solar corona positioned at a height above the limb varying according to the spacecraft heliodistance.

This is achieved by imaging the full off-limb corona in polarized visible light in the band 580-640~nm and in the ultraviolet H~{\sc i} Ly$\alpha$ line emitted by neutral hydrogen at 121.6~nm in a field of view 1.6-$2.9^\circ$ wide, and by deriving from these data acquired at high temporal cadence coronal maps of the speed of the main component of the solar wind -- the proton component -- at a given time.

The UVCS observations of the solar minimum corona proved the interdependence of the geometry of the flux tubes channeling the outflows in the solar atmosphere and the wind velocity as proposed by \citet{wang1990}, who showed that the magnetic field lines divergence inferred at coronal heights and quantified by the fluxtube expansion factor is anticorrelated with the wind speed measured \emph{in situ} at 1~AU.
The UVCS data \citep[see, for instance, the reviews by][]{antonucci2006,cranmer2007,antonucci2012} confirmed the results of the semiempirical models developed and reviewed by \citet{wang2012,wang2020} and MHD models of the corona \citep[e.g.,][]{cohen2015}.
That is, the divergence of the open magnetic field lines rooted in the solar minimum polar coronal hole rapidly increases from the core to the hole boundary and as the areal expansion of the flux tube increases the outflow speed of the coronal plasma decreases.
This implies that during solar minimum the large polar coronal holes are not only the sources of the fast wind, emanating right from their core, but they significantly contribute to the slow wind originating in their peripheral regions and observed at lower heliolatitudes \citep{antonucci2005}. Furthermore, the UVCS observations showed the relevance of magnetic field lines characterized by non-monotonic expansion factors in determining the slowest solar wind component \citep[e.g.,][]{noci1997,antonucci2006}.

The release into the slow wind of coronal plasmoids is considered an additional source of plasma altering the quasi-steady conditions of the slow solar wind originating from coronal holes.
At sunspot minimum, plasma `blobs' are formed by magnetic reconnection processes at the interface between coronal hole field lines and streamer loops and they are well observed at the boundary or close to the cusp of streamers \citep{sheeley1997,wang1998,sheeley2009,wang2012}.
These magnetic features are thus confined to a narrow region around the plasma layer embedding the coronal-heliospheric current sheet, and very likely they provide only a limited fraction of the slow solar wind \citep{wang2020}.
Density structures, formed $\leq 2.5$~$R_\odot$, flowing near streamers with a periodicity of about 90 minutes have also been observed in the inner Heliospheric Imaging data collected with the STEREO/SECCHI suite \citep[][]{viall2015}.  
Structures with similar characteristics are then observed in the heliosphere \citep[][]{kepko2016,kepko2020,viall2021}.
In the same line, \citet{sanchez-diaz2017} find quasi-periodic bursts of activity at the streamer cusps due to intermittent magnetic reconnection associated with the release of small transients.
Recently the coronal origin of magnetic switchbacks, observed \emph{in situ} in the solar wind as abrupt temporary magnetic field reversals, has been identified with Metis as they propagate across the solar corona.
The release of this kind of ejecta is ascribed to the occurrence of interchange reconnection and represents an additional contribution to the slow wind \citep{telloni2022a}.

One further appealing hypothesis on the origin of the slow solar wind has been put forward considering that the MHD models predict the existence of isolated or narrowly connected open field regions in the corona that form a web of quasi separatrix layers, named S-web \citep{antiochos2012,scott2018}.
The photospheric dynamics stresses the separatrix layers inducing current sheets and consequent reconnection processes with release of coronal plasma.
Thus, the outflows generated in the corona by these processes are proposed to provide a significant contribution to the slow solar wind plasma.
Evidence for coronal web structures associable with slow wind streams has been recently presented in a study of ultraviolet observations of the middle corona \citep[][]{chitta2022}.
The observational evidence and hypotheses on the solar wind sources are discussed in several papers over the past decade \citep[e.g.,][]{antiochos2012,antonucci2012,abbo2016,wang2012,wang2020,cranmer2017,antonucci2020a}.

The Metis observations are here analyzed in order to trace the wind evolution in the outer corona, to study the dependence of the wind properties on the magnetic field configuration, and to investigate how the wind plasma is related to the different sources invoked to explain the slow solar wind and its properties which are known to be markedly different from those characterizing the fast wind.
That is, the slow-speed solar wind, flowing in the heliosphere at velocities $\leqslant 450$~km~s$^{-1}$, is characterized by higher temporal and spatial variability, higher ionic charge states and higher abundance of elements with low first ionization potential \citep[e.g.,][]{vonsteiger2000,zurbuchen2002} than the fast wind flowing in quasi-steady conditions at higher velocities $\geqslant 550$~km~s$^{-1}$.
In addition recently, \citet{ko2018} proposed to consider the level of velocity fluctuations observed \emph{in situ} as a robust criterion to classify the wind regimes, since in the slow wind the velocity fluctuations are systematically lower than in the fast one.

In this paper, we analyze the Metis observations of the solar corona obtained on May 15, 2020, during the sunspot minimum of cycle 24 at a height of a few solar radii between 3.8 $R_\odot$ and 7.0 $R_\odot$, in order to derive the fine structure of the solar wind density and velocity as a function of latitude.
These are the observations that provided the first map of the slow solar wind \citep[][hereafter referred as Paper~{\sc i}]{romoli2021}.
On May 15 at the East limb the streamer belt extended outward in the form of a slightly warped, quasi-equatorial layer of denser plasma surrounding the coronal current sheet.
Hence because of this simple quasi-dipolar magnetic configuration, the study of the East limb represents a favorable test case to investigate the coronal sources of the slow wind plasma.

Aims of the present analysis are: first, to study the structure of the slow wind zone in the quiet corona out to the boundary between slow and fast wind, and, second, to discuss the properties of the slow wind in terms of the magnetic field topology, that is, of the expansion of the coronal magnetic field lines and of the physical factor characterizing the separatrices web in the solar atmosphere.
In order to achieve these goals, the analysis of the structure of the coronal wind has been performed in an as detailed as possible way.
For this purpose, a refined approach of the Doppler dimming technique to be applied to the Metis data is introduced, in such a way to take fully into account the best approximation to the actual distribution of the polarized emission and coronal density along the line-of-sight, which is inferred using a 3D magnetohydrodynamic (MHD) model of the solar corona.

\section{\label{sec:obs}Observation of the streamer belt in the solar-minimum corona}

\begin{figure*}
    \centering
    \includegraphics[trim=0.25cm 0.5cm 2.5cm 2cm,clip,width=0.32\textwidth]{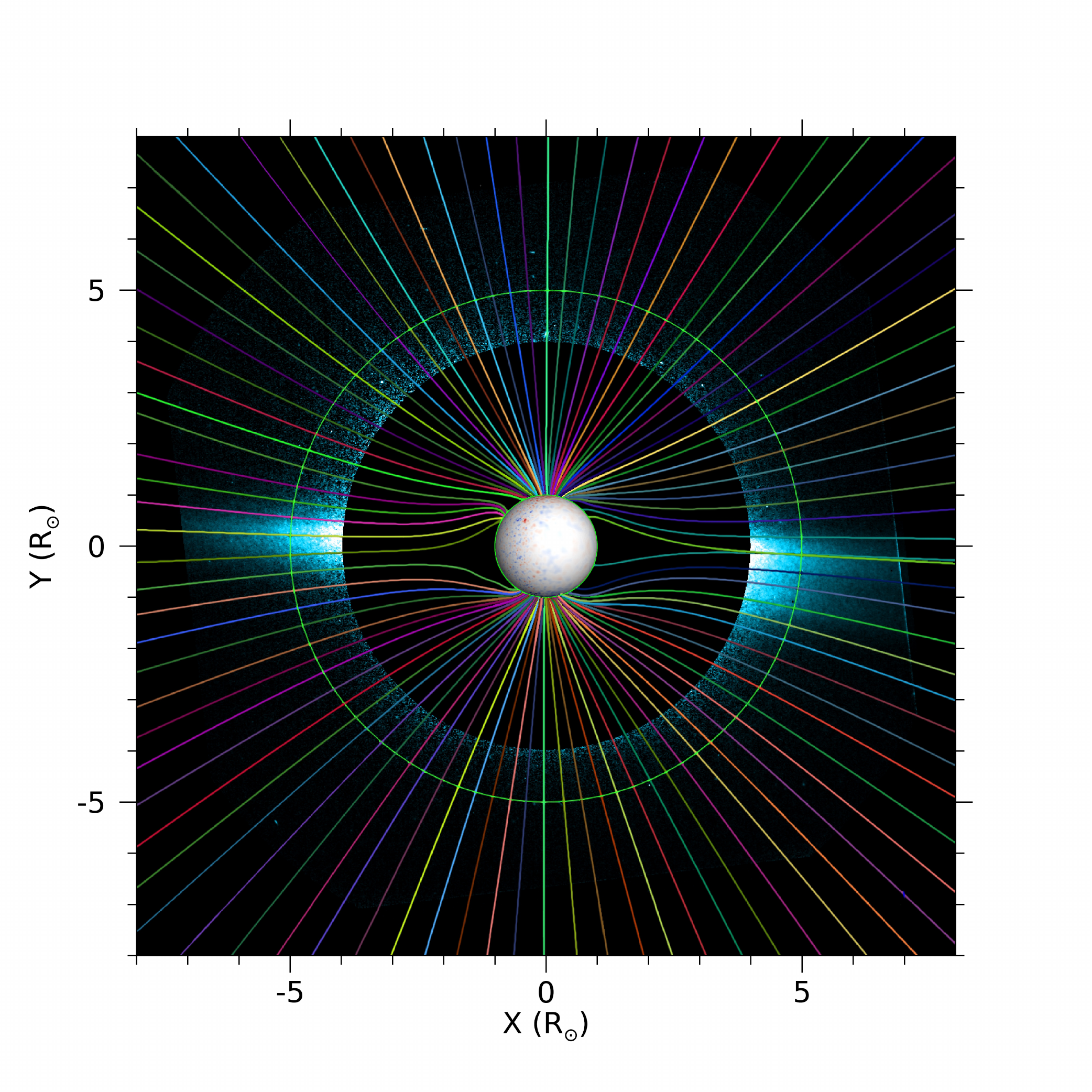}
    \includegraphics[trim=0.25cm 0.5cm 2.5cm 2cm,clip,width=0.32\textwidth]{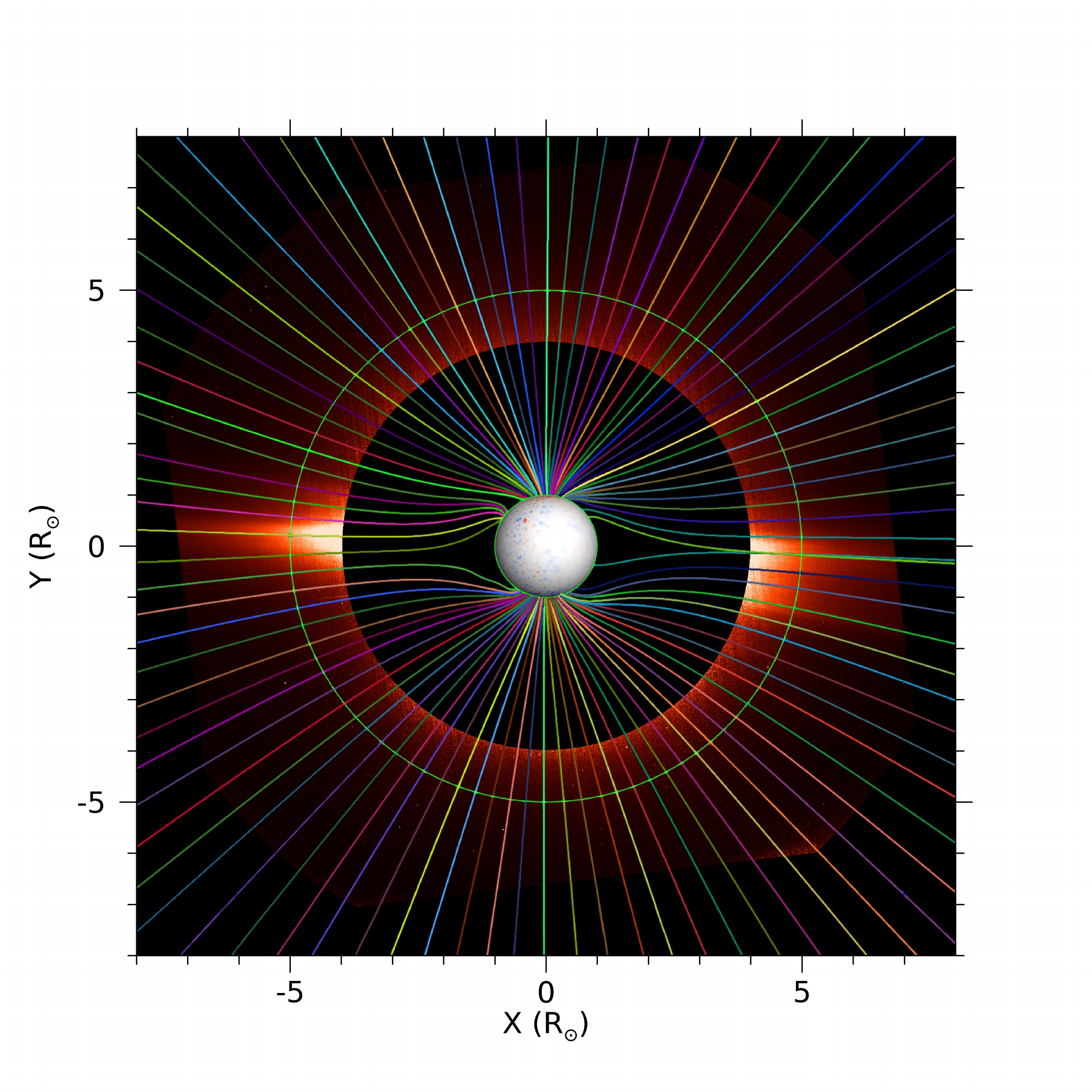}
    \includegraphics[trim=0.25cm 0.5cm 2.5cm 2cm,clip,width=0.32\textwidth]{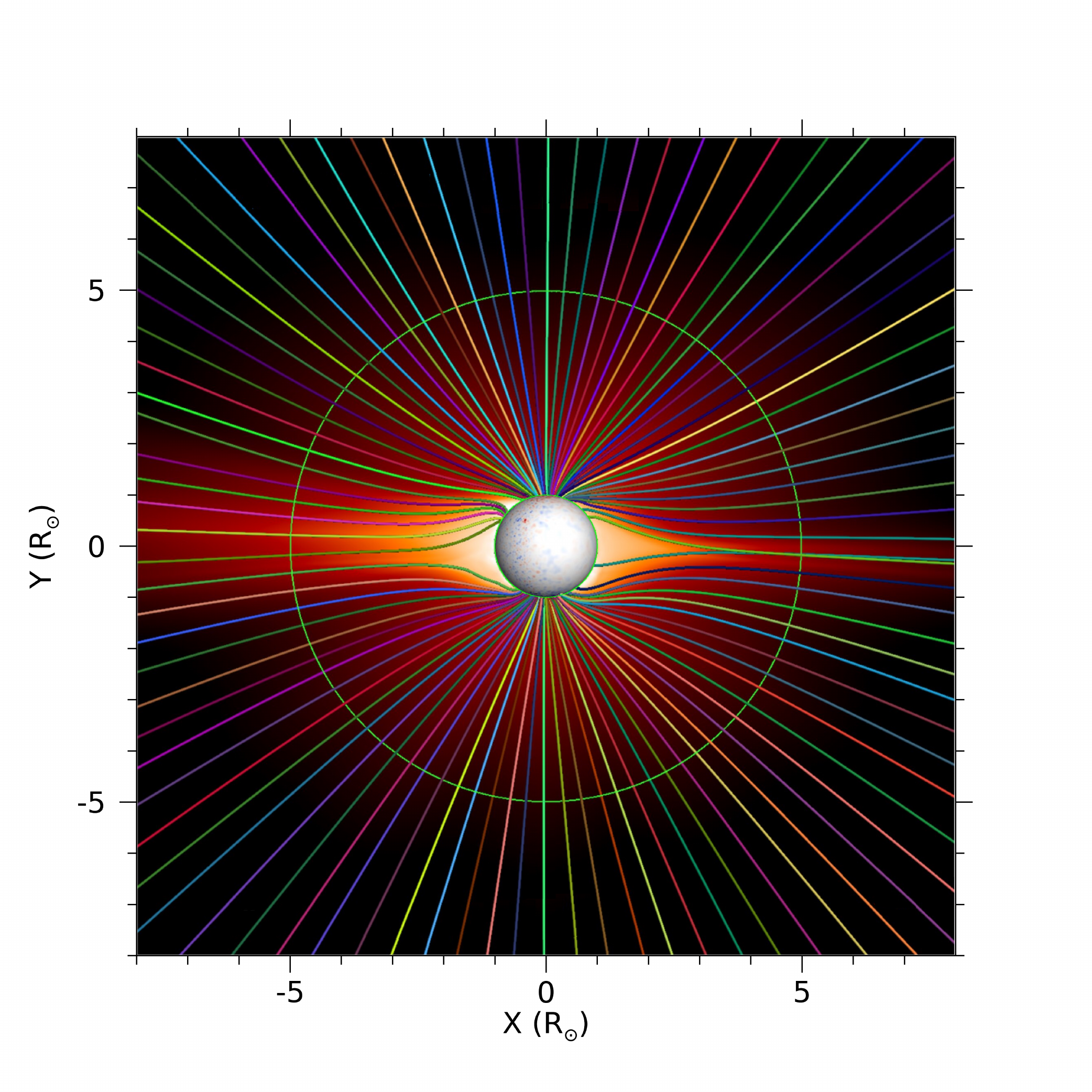}
    \caption{\label{fig:1}Left panel: image of the solar corona H~{\sc i} Ly$\alpha$ intensity detected in the UV channel of Metis, from the vantage point of 0.64~AU on May 15, 2020, at 11:40 UT, in the field of view ranging from 3.8~$R_\odot$ to 7.0~$R_\odot$. The image is overlaid on the magnetic field lines resulting from the PSI 3D MHD simulation traced near the plane of the sky. Middle panel: coronal polarized brightness detected within the visible-light 580-640~nm band of the Metis coronagraph, at the same time and in the same field of view as the left panel. Right panel: composite image of the simulated magnetic field lines and polarized brightness. In all panels, coordinates are given in the Helio-projective Cartesian reference frame and the vertical axis corresponds to the solar rotational axis.}
\end{figure*}

When the Metis coronagraph onboard Solar Orbiter imaged the solar corona for the first time on May 15, 2020 (Figure~\ref{fig:1}, see also Paper~{\sc i}), the quiet equatorial streamer belt characteristic of the sunspot minimum was well observed only at the East limb, whilst close to the Metis observation time a small flux rope was released along the streamer at the West limb (Wang Y.-M., private communication).
The East limb corona is thus suitable for studying the super-radial expansion of the magnetic field lines as a modulator of the solar wind speed at and in the vicinity of the outward extension of the streamer belt where a layer of denser plasma embeds the current sheet, as well as for relating the slow wind belt to the separatrices web region.
For statistical reasons the study is performed in a region about $\pm30^\circ$ wide in latitude across the solar equator.

The region selected for the analysis corresponds to the layer where the coronal slow wind is expected to flow during the solar activity minimum, according to the present knowledge based on in situ and coronal observations.
The \emph{in situ} observations acquired during the first orbit with the Solar Wind Ion Composition Spectrometer, SWICS/Ulysses \citep{gloeckler1992}, show that in the heliosphere the slow solar wind is restricted to a helio-latitude range of $\pm21^\circ$ centered on the streamer belt.
In addition, when the monitoring can be carried on continuously through a solar rotation, the slow wind is found to be confined in an even narrower region around the equator, $13^\circ$ wide. While, the fast wind is observed only outside the $\pm29^\circ$ latitude range \citep{woch1997}.
When Ulysses was crossing the northern hemisphere, the coronal outflows were monitored with UVCS.
\citet{dolei2018} in an analysis of the UVCS data collected in June 1997 found that the slow wind observed at mid-low latitudes is separated by rather well-defined boundaries, at approximately $\pm30^\circ$ from the equator, from the fast wind emerging from the core of the polar coronal holes with the speed of the proton component approximately equal to 300~km~s$^{-1}$ between 2.5~$R_\odot$ and 3.5~$R_\odot$.
Based on an extended study of the oxygen ion component of the coronal wind detected with the UVCS data from May 1996 to August 1997, \citet{telloni2022b} identify a slow/fast wind boundary at $\pm35^\circ$ by studying the correlation of the oxygen kinetic temperature and velocity that is ascribed to a possible development of Kelvin-Helmoltz instability.
The latitudinal extent of the zones characterized by the slow wind is confirmed by the results obtained by \citet{tokumaro2010} from the radio scintillation observations performed over two solar cycles at heliocentric distances above 20~$R_\odot$.
These authors conclude that during solar minima the fast solar wind is strongly predominant in the high-latitude regions $>\left|70^\circ\right|$ and the slowest solar wind flows within $\pm10^\circ$ from the magnetic neutral line.
However as can be deduced from the velocity maps derived from the radio scintillation data, the slow wind fills a wider region encompassing the current sheet.
The study of the heavy ion composition, measured with the Advanced Composition Explorer, ACE, and used to determine the origin of the solar wind at the Sun, leads to the conclusion that the sources of `non-coronal-hole' wind are within a region of average width $<17^\circ$ during the anomalous solar minimum of cycle 23, smaller than that observed at the end of the previous solar cycle \citep{zhao2009}.
The differences in the characteristics of two successive solar minima point to a possible 22-year cycle dependence of the coronal magnetic field configuration.

In the coronal region selected for the slow solar wind analysis, the outflow velocity of the plasma is derived on the basis of the simultaneous Metis observations in the visible-light (VL) and ultraviolet (UV) channels by adopting the most likely distribution of the electron density along the line of sight (LOS) in order to account for the moderate warping of the equatorial plasma sheet.
The velocity results are then discussed in terms of the expansion factor, $f$, measuring the divergence of the magnetic field lines and the squashing factor, $Q$, which is a topological measure of the separatrices and quasi-separatrix layers \citep{titov2002,titov2007}, and allows to identify the zone covered by the separatrix web \citep{antiochos2012}.
The parameters related to the coronal magnetic field topology as well as the functional dependence of the distribution of the electron density along the line of sight, are inferred from a 3D MHD model of the solar corona based on SDO/HMI synoptic map data \citep{scherrer2012}, as described in Sect.~\ref{sec:dimming}.

\subsection{\label{sec:metis_obs}Metis observations on May 15, 2020}
The first Metis image of the solar corona was obtained during the Solar Orbiter commissioning phase on May 15, 2020, in the time interval 11:39-11:41~UT when the coronagraph was observing the Sun at a distance of 0.64~AU and at an angle of $11.4^\circ$ West relative to the Earth-Sun direction.
The spacecraft was at $4.3^\circ$ North relative to the solar equatorial plane.
The solar corona was observed in an annular region between 3.8~$R_\odot$ and 7.0~$R_\odot$ in visible light polarized brightness, $pB$, in the wavelength band ranging from 580~nm to 640~nm, and in the ultraviolet band $121.6\pm10$~nm where the H~{\sc i} Ly$\alpha$ light is emitted.
The acquisitions used in the analysis are: the $pB$ sequence consisting of four polarimetric frames with detector integration time equal to 30~s for the VL channel, and one set of 6 UV images, averaged over 6 frames each with integration time of 16~s for the UV channel.
The VL and UV detectors were configured to obtain an image scale on the plane of the sky of about 4300~km and 8600~km, respectively.
In this analysis we consider the region within $\pm30^\circ$ from the equator at the East limb where the UV signal-to-noise (S/N) ratio is $\gtrsim 3$.

The simultaneous detection of the polarized brightness $pB$ in the VL channel and the H~{\sc i} Ly$\alpha$ emission in the UV channel allows a measure of the electron density and the outflow velocity of the coronal wind in the coronagraph field of view.
The polarized brightness is due to the scattering of the photospheric light by the free electrons of the K-corona.
The observed H~{\sc i} Ly$\alpha$ line is mainly emitted by the residual neutral hydrogen atoms present in the hot corona through resonant scattering of the H~{\sc i} Ly$\alpha$ exciting radiation coming from the chromosphere \citep[][]{gabriel1971}.
The emission is subject to the dimming caused by the Doppler shift between the incident chromospheric photons and the spectral absorption profile of the coronal hydrogen atoms moving outward with the wind flow.
In the frame of reference of the plasma outflows, the spectrum originating in the chromosphere appears to be red-shifted and the radiative excitation rate depends on the wind speed.
The outflow velocity is thus derived by calculating the Doppler dimming of the resonantly scattered UV emission relatively to the value expected for a static corona when the adopted plasma electron density is that measured with the polarized visible light data.
The Doppler dimming diagnostics \citep[][]{beckers1974} for deriving the wind speed in the corona has been introduced by \citet{withbroe1982} and \citet{noci1987}, and for what concerns the hydrogen component of the wind this technique is applicable in the outflow velocity range approximately from 100 to 350~km~s$^{-1}$.
The high rate of charge exchange between protons and neutral hydrogen at the coronal heights under study ensures that the neutral hydrogen can be used as a proxy for the protons \citep{withbroe1982,allen1998}.

The data are calibrated according to the procedures discussed in Paper~{\sc i}, except for what concerns the UV radiometric calibration, which has been improved by considering the detection of a larger set of standard UV stars transiting across the Metis field of view.
The stars selected for this purpose are, in particular, $\alpha$ Leonis, $\rho$ Leonis, observed on June 15 and 17, 2020, and $\omega$ Scorpii and $\theta$ Ophiuchi, on March 15 and 25, 2021, respectively.
Their transits across Metis field of view occurred in the coronal region approximately between $\pm30^\circ$ in heliographic latitude, thus the measurement of their UV fluxes is suitable to better characterize the region of interest in the present analysis.
Fluxes measured along the different stellar trajectories, which in the region surrounding the solar equator are more radially aligned, have been interpolated in the radial direction and averaged over specific latitudinal ranges in order to derive a 2D radiometric efficiency map covering the full range of heights in the Metis FOV, thus taking into account the radial dependence of Metis optical vignetting function and possible non-uniformities in the radiometric response of the UV channel within the field of view above the East limb.
A more comprehensive analysis considering all the UV stars observed by Metis is going to be finalized (De Leo et al., in preparation), providing an average radiometric efficiency map to be used to calibrate the full instrument field of view.
Ly$\alpha$ intensities resulting from the \emph{ad hoc} calibration procedure adopted for the present work and the general one considering the full set of UV stars are however consistent within uncertainties $\lesssim 20\%$ on average.
The adopted UV radiometric calibration results in H~{\sc i} Ly$\alpha$  intensities higher than those obtained in Paper~{\sc i}, particularly in the inner coronal regions below 5~$R_\odot$.
This implies lower values for the outflow velocity determined by the Doppler dimming technique and as a consequence a significant outward acceleration of the wind along the central axis of the considered streamer.

\section{\label{sec:dimming}Doppler-dimming analysis of the H~{\sc i} Ly$\alpha$ emission}
The electron density and the speed of the hydrogen atom outflows in the corona are derived from the $pB$ data and the Doppler dimmed H~{\sc i} Ly$\alpha$ emission, respectively, with a refined version of the analysis procedure discussed in Paper~{\sc i}, where the electron density used to compute the H~{\sc i} Ly$\alpha$ emission expected for a static corona has been obtained from the Metis coronal $pB$ measurements via the inversion method described in \citet{vandehulst1950}, assuming a cylindrical approximation for the longitudinal density distribution.

\begin{figure*}
    \centering
    \includegraphics[trim=-0.5cm -0.5cm -0.5cm -0.5cm,clip,width=0.4\textwidth]{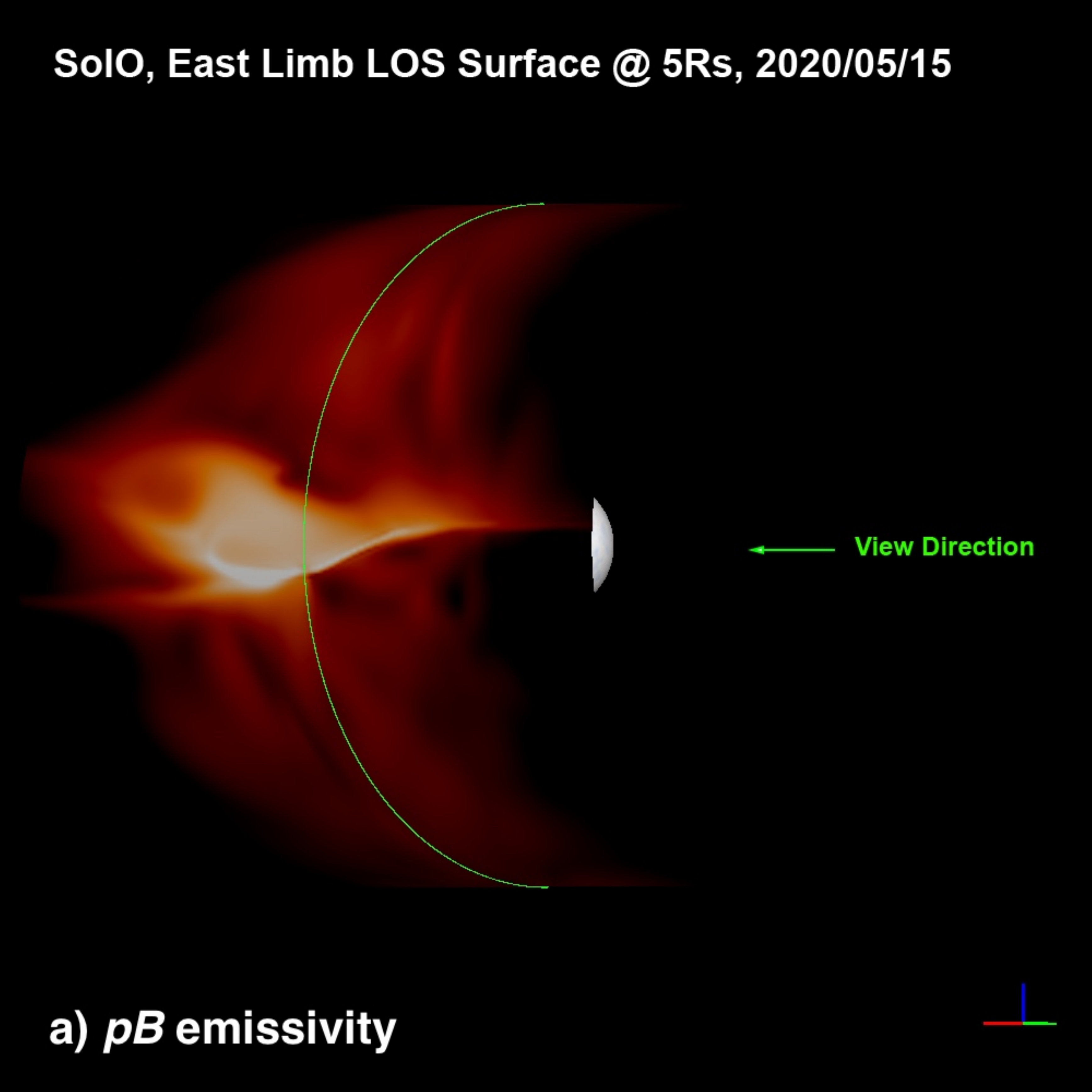}
    \includegraphics[trim=-0.5cm -0.5cm -0.5cm -0.5cm,clip,width=0.4\textwidth]{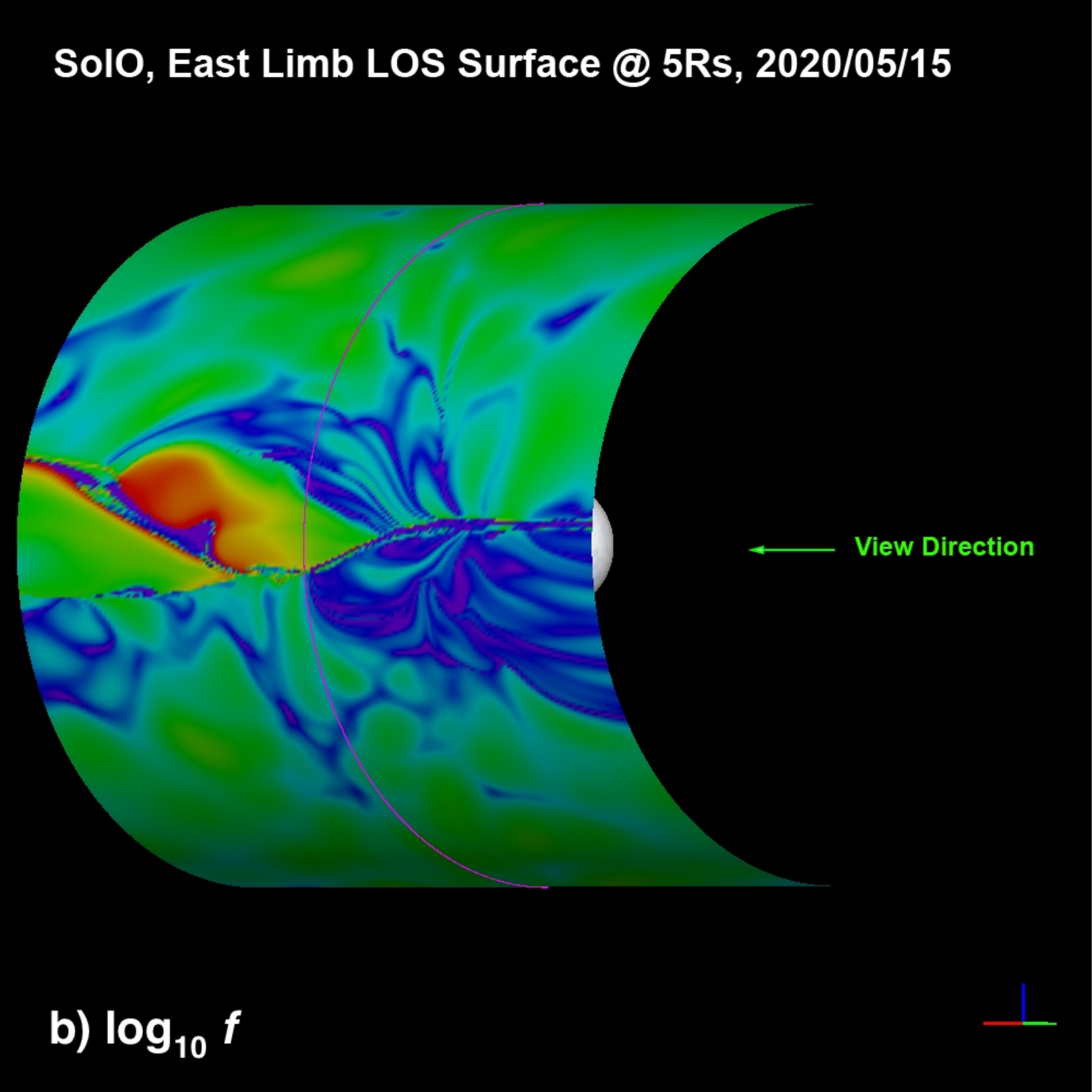}
    \includegraphics[trim=-0.5cm -0.5cm -0.5cm -0.5cm,clip,width=0.4\textwidth]{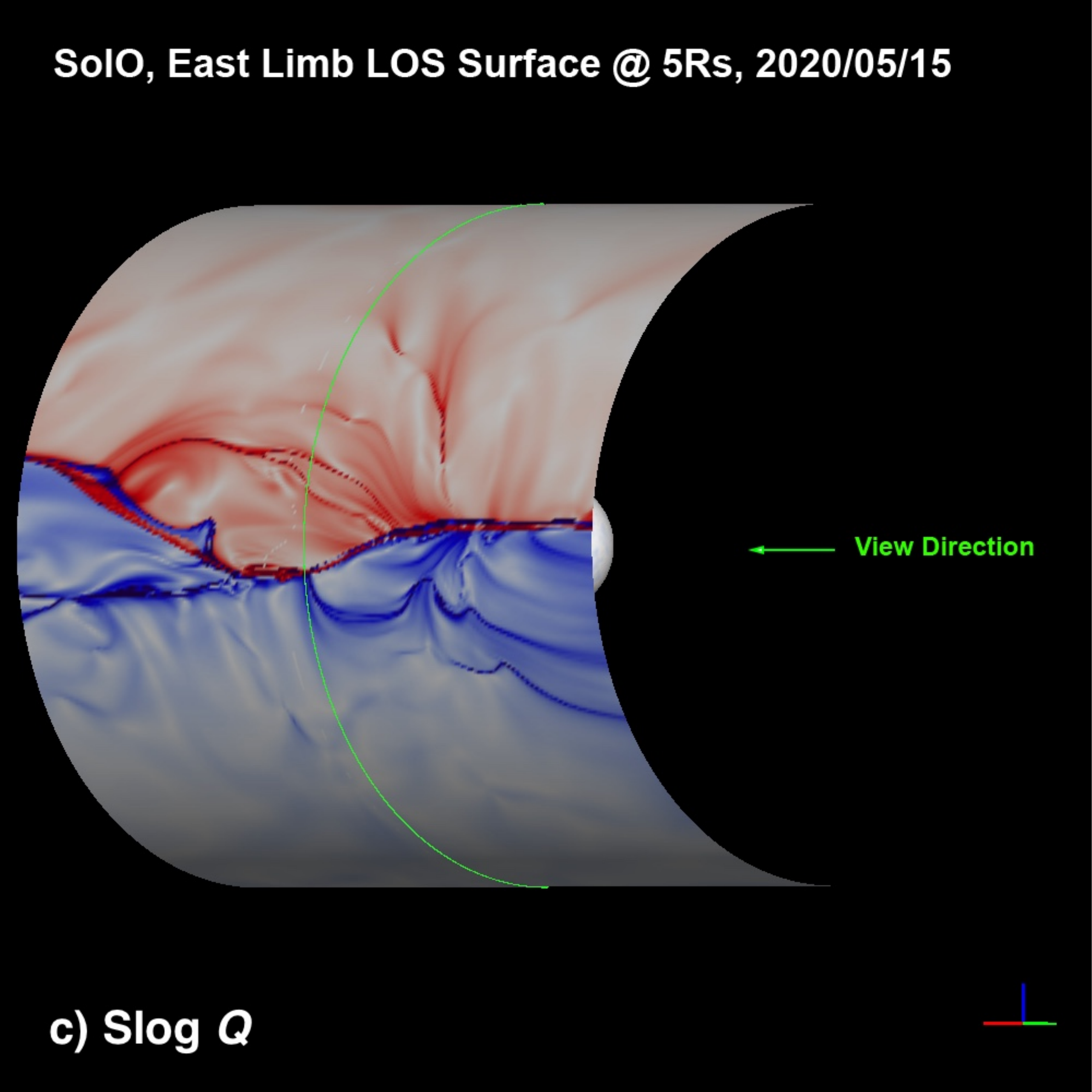}
    \includegraphics[trim=-0.5cm -0.5cm -0.5cm -0.5cm,clip,width=0.4\textwidth]{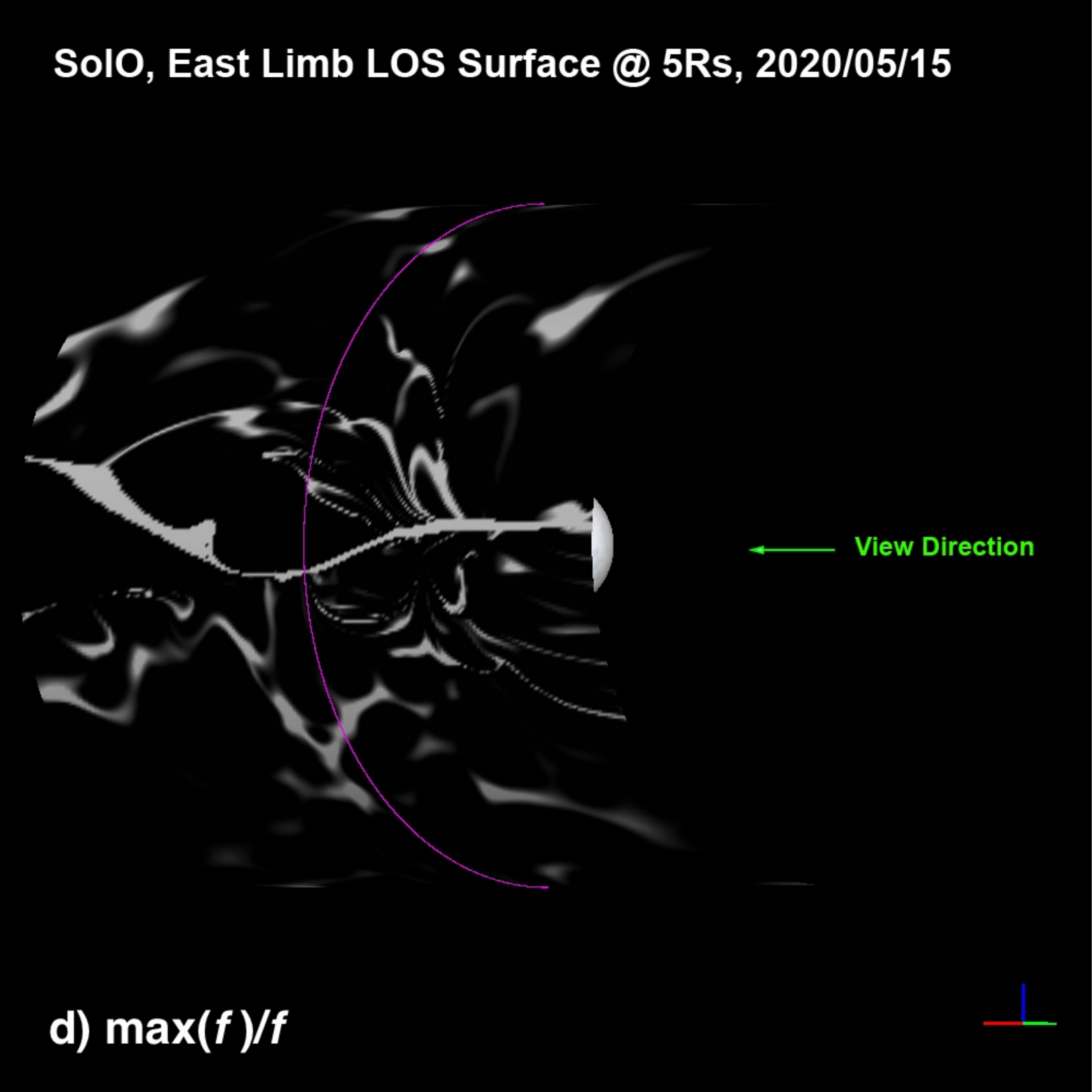}
    \caption{\label{fig:2}The first image a) shows the polarized emissivity (indicated as $pB$ emissivity), calculated with the 3D MAS WTD MHD model, on a cylinder that describes the LOS into/out of the plane of the sky at 5~$R_\odot$ to illustrate the structures present along the LOS at this height. In order to show the cylinder in the East limb view, the viewpoint is rotated of $45^\circ$ in longitude from the Solar Orbiter position, and the direction is indicated with the arrow. The field of view is $\pm8$~$R_\odot$. The LOS dependences of the expansion factor, $\log_{10}\,f$, the signed log of the squashing factor, $\mathrm{slog}\,Q$, and the ratio of the maximum expansion factor relative to the expansion factor calculated at 5~$R_\odot$, $\max(f)/f$, are imaged in the panels b), c), and d), respectively.}
\end{figure*}

In order to resolve the structure of the solar wind outflows in the corona at a finer level, in this study a more detailed description of the corona is adopted.
That is, the functional dependence along the LOS of the polarized emissivity and the electron density taken into account is that returned by a 3D MHD simulation of the corona considering the position of Solar Orbiter with respect to the Sun at the time of the Metis observations.
In Figure~\ref{fig:2}a the LOS surface of the simulated polarized emissivity, which is used to compute the $pB$, highlights the slightly warped equatorial plasma/current sheet that is outlining the extension of the streamer belt in the outer corona.
The magnetohydrodynamic model of the global corona used here is the thermodynamic MHD model, MAS, developed by Predictive Science Inc. \citep{lionello2009,mikic2018}.
This model uses a wave-turbulence-driven (WTD) approach to specify coronal heating (see \cite{mikic2018} for a description of the 3D implementation and \cite{boe2021,boe2022} for a benchmark of the exact WTD model parameterization used here).
In the present study, the choice of the full-Sun magnetic data used to model the corona is optimized to better reflect the conditions at the time of the Metis East limb observations; for this reason, the West limb conditions are not represented at best as shown in Figure 1.
The data chosen to define the boundary conditions include the synoptic maps for the Carrington rotations CRs 2230 and 2231 as well as the daily magnetogram taken about $50^\circ$ away from disk center on May 20, 2020, to characterize the magnetic field in the regions lying to the East of the limb considered in the analysis.

\begin{figure}
    \centering
    \includegraphics[trim=0 0 0 2cm,clip,width=\columnwidth]{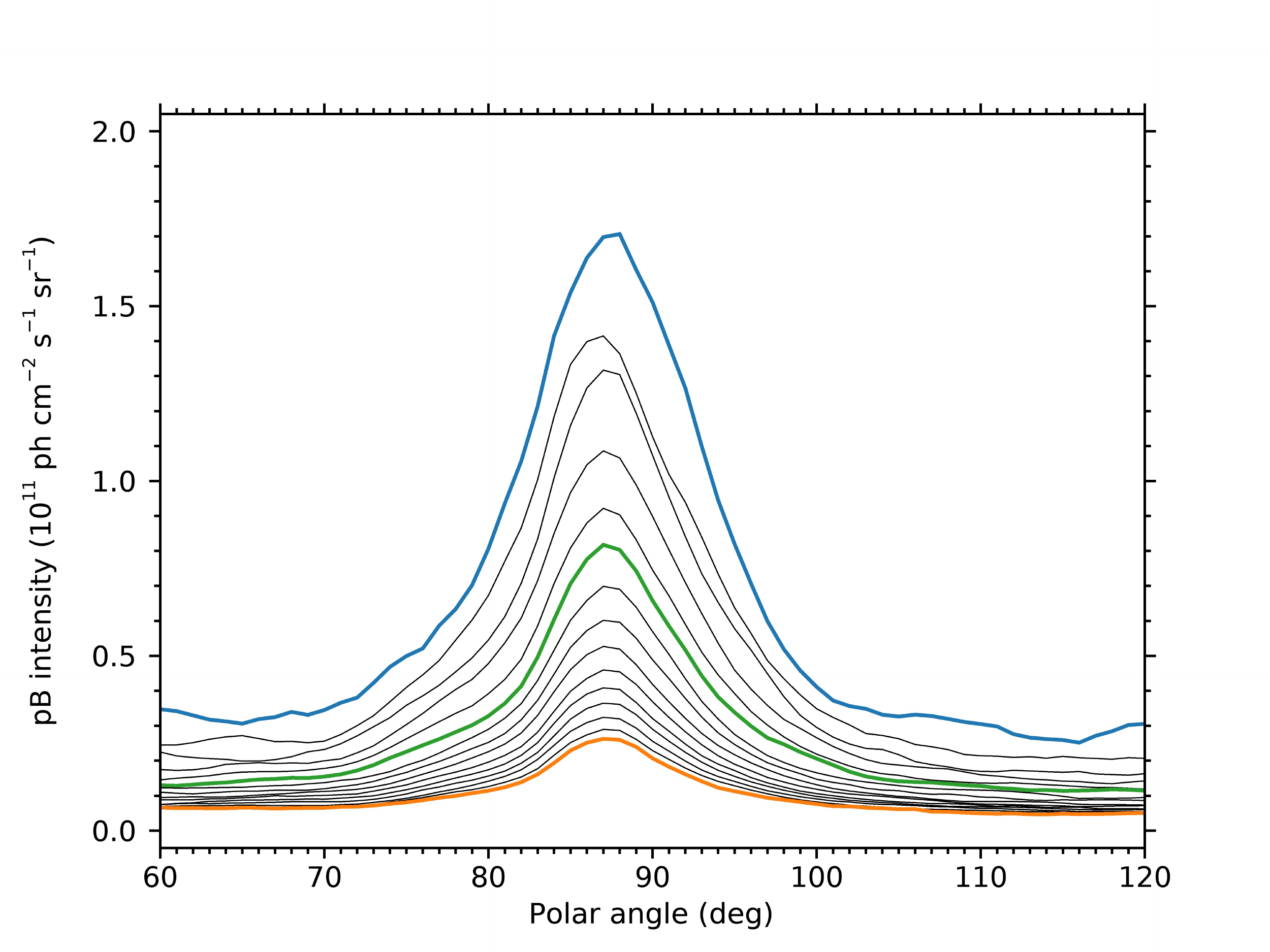}
    \includegraphics[trim=0 0 0 2cm,clip,width=\columnwidth]{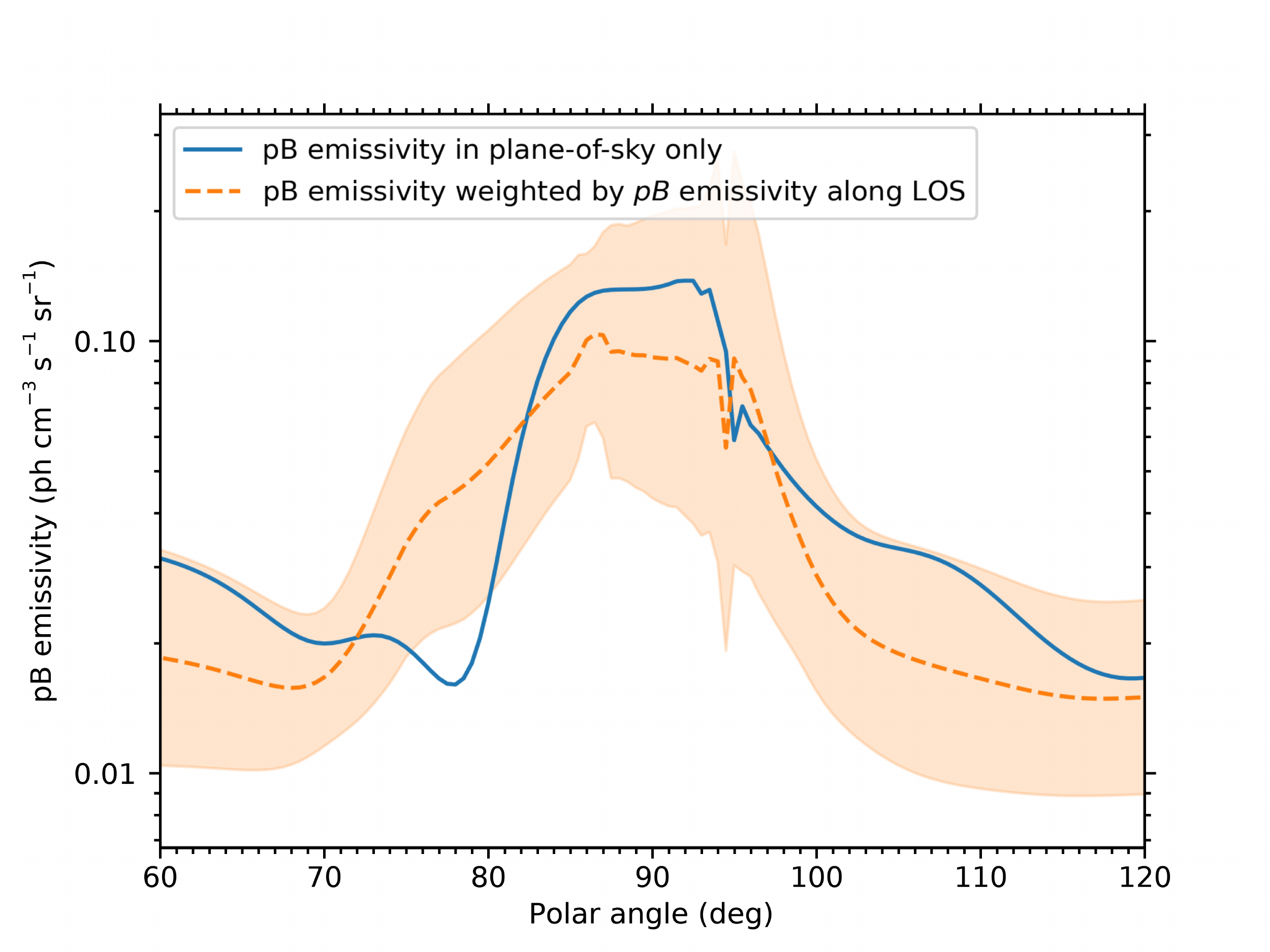}
    \caption{\label{fig:3}Upper panel: latitudinal profiles at the East limb of the polarized brightness (indicated as $pB$) intensity, derived from the VL Metis observations from 4.0~$R_\odot$ to 6.8~$R_\odot$. The latitudinal profiles are reported in steps of 0.2~$R_\odot$ (blue line at 4.0~$R_\odot$, green line at 5.0~$R_\odot$, orange line at 6.8~$R_\odot$). Lower panel: polarized emissivity (indicated as $pB$ emissivity) at 5~$R_\odot$ in the plane of the sky (continuous blue line) and the same quantity averaged along the line of sight and weighted by the polarized emissivity (dashed orange line), calculated by the 3D MHD simulation of the corona. The shaded region indicates the standard deviation ($\pm1$ standard deviation) of the weighted emissivity along the LOS.}
\end{figure}

Figure~\ref{fig:2}a shows the LOS surface of the modelled polarized emissivity in the extended corona on a cylinder that describes the LOS into/out of the plane of the sky at 5~$R_\odot$ at the East limb, to illustrate the plasma longitudinal distribution at this height.
Figure~\ref{fig:3} shows the comparison of the East limb latitudinal profiles of the polarized brightness $pB$ (obtained by integrating the emissivity along the LOS) derived from the Metis VL channel observations from 4.0~$R_\odot$ to 6.8~$R_\odot$ in steps of 0.2~$R_\odot$ (upper panel), with the polarized emissivity in the plane of the sky, POS, at 5~$R_\odot$ - plotted with the same quantity averaged along the line of sight and weighted by the polarized emissivity -, calculated by means of the MHD simulation of the corona (lower panel).
The weighted polarized emissivity accounts for the presence of structures present at the limb with a large longitudinal extension, structures out of the plane of the sky, or multiple structures contributing to the observables.

The functional dependence along the LOS of the model electron density is taken into account in the following way.
The values of the model electron densities along each selected LOS direction are normalized by multiplying by a factor obtained from the ratio between the $pB$ measured by Metis and that derived from the model polarized emissivity integrated along the LOS.
This factor ranges from 0.9 to~1.5 in the selected latitudinal range at the East limb, and is nearly constant along each radial direction considered within this range.
The model polarized emissivity is based on the local electron density from the 3D MHD model and the polarized emissivity kernel, which depends on the scattering geometry between the Sun and the observer for the plasma at each point along the LOS \citep[for example, see][]{howard2009}.
The range over which the integration is performed in order to get the modeled $pB$ depends on the heliocentric distance on the plane of the sky.
In fact, the data cube returned by the MHD simulation has a longitudinal extension of $\pm60^\circ$ from the POS.
Therefore, the modeled $pB$ for a given point on the POS at a distance r from the Sun has been obtained by integrating the modeled polarized emissivities along a total length with half-width (HW) given by
\begin{equation}
    LOS_{HW}(r)=r\cdot\tan (60^\circ)=r\cdot\sqrt{3}.
\end{equation}
For instance, the minimum and maximum values of the LOS half-width are 6.6~$R_\odot$ and 12.1~$R_\odot$ for $r=3.8$~$R_\odot$ and $r=7.0$~$R_\odot$ on the POS, respectively.

At this point, the synthetic coronal H~{\sc i} Ly$\alpha$ line intensity at a fixed point on the POS can be calculated by integrating the emissivity along the same variable LOS ranges used for the integration of the polarized emissivity.

\subsection{\label{sec:dimming_assumption}Doppler-dimming analysis assumptions}
The input quantities for the Doppler dimming analysis are: the observed Doppler dimmed intensity of the H~{\sc i} Ly$\alpha$ line obtained from the Metis data, corrected for the interplanetary H~{\sc i} Ly$\alpha$ intensity, and the intensity values estimated in the assumption of a static corona.
The physical parameters needed as input to perform the Doppler dimming calculations are the electron temperature, the hydrogen ionization fraction, the helium abundance, the exciting chromospheric H~{\sc i} Ly$\alpha$ line intensity and profile, and the hydrogen kinetic temperature.

The adopted interplanetary H~{\sc i} Ly$\alpha$ intensity is equal to $3.0\times10^7$~photons~s$^{-1}$~cm$^{-2}$~sr$^{-1}$ \citep{kohl1997,suleiman1999}.
The electron temperature, $T_e$, is assumed constant with latitude, within $\pm30^\circ$, and varies with heliodistance according to the extrapolation of the temperature curve given by \citet{gibson1999}.
The hydrogen ionization fraction is determined on the basis of the electron temperature values.
The helium abundance is assumed to range from 2.5\% (equatorial latitude) to 10\% at the boundaries of the considered latitudinal range, according to the observations obtained by \citet{moses2020} during the 2009 solar minimum.
In the alternative hypothesis of a uniform value of $10\%$ of the helium abundance over the latitude range analyzed the speed derived at 6.8~$R_\odot$ varies of $\sim3\%$.
For what concerns the H~{\sc i} Ly$\alpha$ chromospheric line, its intensity is assumed constant and equal to $8.16\times10^4$~erg~cm$^{-2}$~s$^{-1}$~sr$^{-1}$ ($5.0\times10^{15}$~photon~cm$^{-2}$~s$^{-1}$~sr$^{-1}$), as in Paper~{\sc i}.
The line profile is represented by the analytic function reported by \citet{auchere2005} and is assumed to be uniform over the solar disk, since it is proved that the non-uniformity of the shape of the H~{\sc i} Ly$\alpha$ line profile observed in the chromosphere does not induce remarkable effects on the estimate of the outflow velocity \citep{capuano2021}.

The most critical physical parameter to be adopted in the computation of the outflow velocity is the kinetic temperature, $T_k$, of the hydrogen atoms, that, depending on the assumptions, leads to a difference in the estimate of the velocity value from few~km~s$^{-1}$ at 4~$R_\odot$ to ~ 40~km~s$^{-1}$ at 6.8~$R_\odot$.
The assumptions on $T_k$ and its degree of anisotropy thus significantly affect the outflow velocity derived by applying the Doppler dimming analysis.
The UVCS observations have shown that the velocity distribution of the hydrogen atoms and ions, which is due to the combined effects of thermal motions, non-thermal motions and outflow of coronal plasma into the solar wind, turns out to be bi-Maxwellian at least in the polar regions of the solar minimum corona.
The kinetic temperature, $T_{k,\perp}$, is directly derived from the width of observed spectral lines and measures the spread of the velocity distribution along the LOS, that is, perpendicularly to the radial magnetic field lines.
This same quantity, however, cannot be directly measured across the LOS, thus the value of $T_{k,\parallel}$ is to some extent uncertain.
Constraints on the degree of the kinetic temperature anisotropy, $T_{k,\perp}/T_{k,\parallel}$, can be put by applying the Doppler dimming analysis.

In the polar regions observed at solar minimum and in the case of lines emitted by heavier ions, such as the five-time ionized oxygen, the ion velocity distribution turns out to be bi-Maxwellian with the ion velocity distribution much broader across than along the magnetic field lines, $T_{k,\perp}\gg T_{k,\parallel}$.
This strong anisotropy has been interpreted as the effect of preferential energy deposition across the magnetic field in the fast wind, as the result of damping of transverse waves due to ion-cyclotron resonance \citep{kohl1998,cranmer1999a,cranmer1999b}.
The degree of kinetic temperature anisotropy of the oxygen component still compatible with the data increases to a peak value at about 2.9~$R_\odot$ ($T_{k,\perp}/T_{k,\parallel}\simeq 14$, lower limit) and then decreases toward less anisotropic conditions around 3.7~$R_\odot$ according to \citet{telloni2007b}.
Discussion on the five-time ionized oxygen observations in a series of papers \citep[e.g.,][]{raouafi2006}, however, suggests that isothermal conditions for heavy ions, protons and electrons cannot be completely ruled out.

For what concerns the hydrogen atoms, the Doppler-dimming analysis of the H~{\sc i} Ly$\alpha$ lines does not provide strong constraints on the degree of the kinetic temperature anisotropy as it does for the lines emitted by heavier ions.
The first analyses of the H~{\sc i} Ly$\alpha$ lines indicate higher proton temperatures, 3.8~MK at 4~$R_\odot$, perpendicular to the magnetic field \citep{cranmer1999a}.
A more recent analysis by \citet{cranmer2020}, taking into account the UVCS measurements of the H~{\sc i} Ly$\alpha$ lines in polar coronal holes out to 4~$R_\odot$ and a large ensemble of empirical model results, concludes that there is evidence for weak dissipation of Alfvén waves increasing with height, with typical values for parallel and perpendicular proton temperatures which are not much different, 1.8~MK and 1.9~MK, respectively.
In addition, these values do not exhibit much variation between 1.4~$R_\odot$ and 4~$R_\odot$. In any case, in the denser low latitude equatorial regions, the conditions of isotropic velocity distribution and thermal equilibrium between species are more likely reached than in the polar holes \citep[e.g.,][]{vasquez2003}.

The value of the H~{\sc i} kinetic temperature $T_{k,\perp}=1.6$~MK along the LOS, assumed in this study, is derived from a large database of H~{\sc i} Ly$\alpha$ line profiles obtained with UVCS during solar minimum \citep{dolei2019}.
This value is also considered to be uniform over the region of interest.
In the present analysis both cases are considered: 1) isotropy ($T_{k,\perp}=T_{k,\parallel}$) over the entire region of interest $\pm30^\circ$ wide around the equator, and 2) anisotropy ($T_{k,\perp}>T_{k,\parallel}=T_e$), with protons in thermal equilibrium with electrons across the line of sight.

\subsection{\label{sec:technique}Diagnostic technique}
\begin{figure}
    \centering
    \includegraphics[trim=0 0 0 2cm,clip,width=\columnwidth]{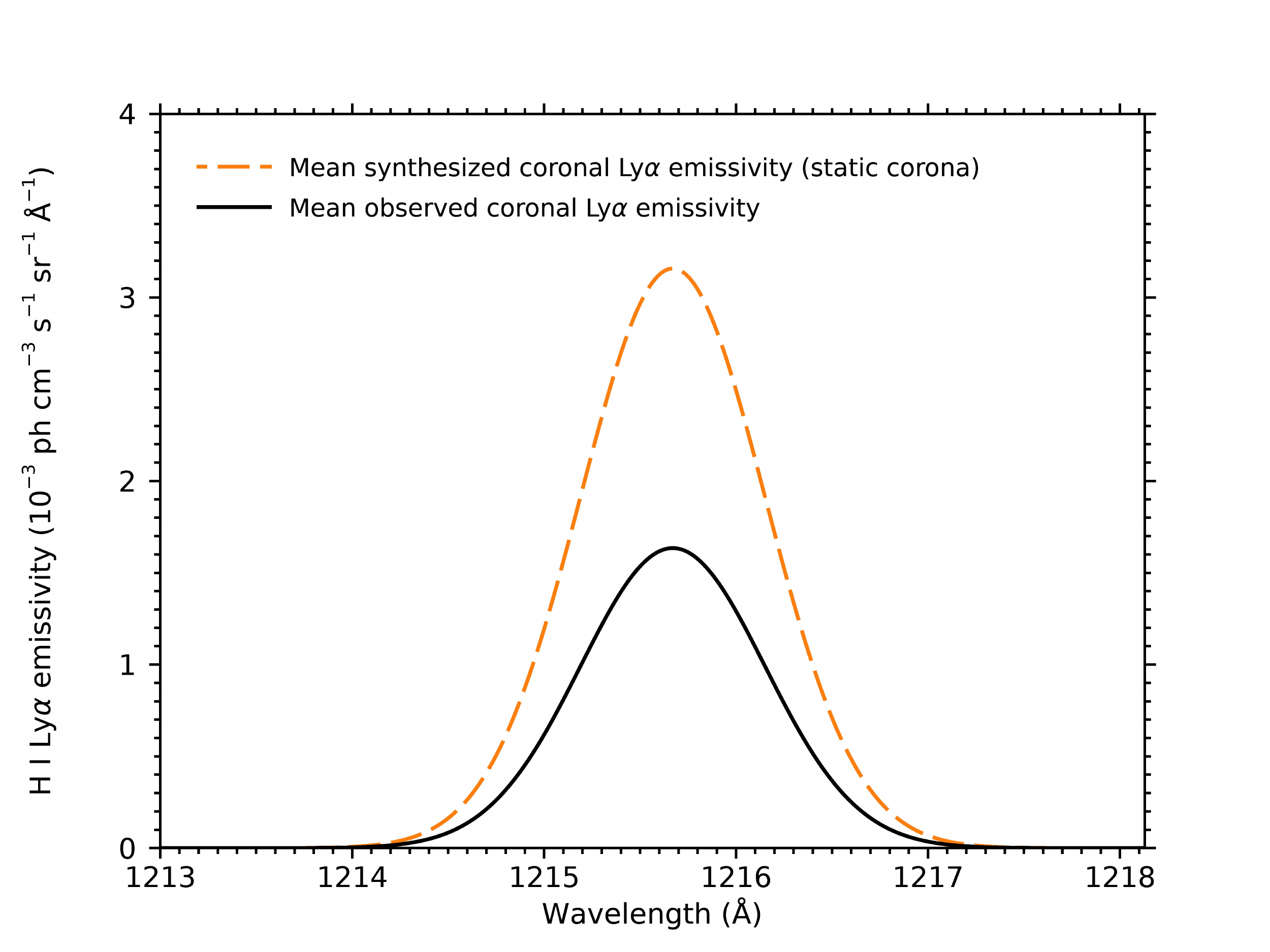}
    \caption{\label{fig:4}Profiles of the H~{\sc i} Ly$\alpha$ spectral line with intensity measured at 5~$R_\odot$ at the East limb on May 15, 2020, in the UV channel of Metis and width consistent with a kinetic temperature of 1.6~MK (continuous line) and of the synthetic H~{\sc i} Ly$\alpha$ line for a static corona, calculated on the basis of the coronal density derived from the observations performed in the Metis VL channel (dashed line).}
\end{figure}

Starting from the physical parameters discussed in the previous sections as inputs, a synthetic value of the coronal H~{\sc i} Ly$\alpha$ emissivity can be calculated in a given point in space.
The synthetic H~{\sc i} Ly$\alpha$ intensity is then obtained by integrating the Ly$\alpha$ emissivity calculated for all points along a given LOS.
The H~{\sc i} Ly$\alpha$ spectral line calculated only on the basis of the density in the hypothesis of a corona in static conditions and the H~{\sc i} Ly$\alpha$ line with intensity as measured in the Metis UV channel, which is dimmed in the presence of wind outflows, are represented in Figure~\ref{fig:4} with a line broadening consistent with a neutral hydrogen kinetic temperature of 1.6~MK.
The comparison of the emission calculated for a static corona and the observed emission -- previously corrected for the interplanetary H~{\sc i} Ly$\alpha$ contribution that is not negligible relative to the coronal values beyond $\sim 5$~$R_\odot$ -- is used to derive the expansion velocity rate in the corona.
The intensity depends on the only free parameter, the outflow velocity, and due to the Doppler dimming decreases as the velocity increases.
The outflow velocity value is returned iteratively until a satisfactory match between the calculated and observed intensities is reached.
All the details concerning the numerical tools and the outflow velocity determination algorithms are reported in \citet{dolei2019,antonucci2020b,capuano2021,telloni2021}.
The Doppler-dimming code used in Paper~{\sc i} has been further validated by comparing the results obtained with those derived with the code by \citet{telloni2021}.

Before applying the Doppler dimming diagnostics, the UV image is converted from rectangular to polar coordinates.
The selected sector at the East limb in polar angle goes from $60^\circ$ to $120^\circ$, counterclockwise from the North solar pole (that is $\pm30^\circ$ with respect to the equatorial plane).
In order to improve the UV image statistics and increase the signal-to-noise ratio, the calibrated UV images are further binned over $4\times4$~pixels, corresponding to a polar angle resolution equal to 1~degree.
Finally, a mean over the polar angle with a step of 3~degrees and over each radial profile with a step of 0.2~$R_\odot$ was performed to further improve the S/N ratio.

\section{\label{sec:results}Results}
The wind outflow velocity and density, detected near the Sun at the East limb on May 15, 2020 in the coronal zone enclosing the current sheet, are reported in the next sections and discussed in the context of the coronal magnetic field topology, which can be characterized by the divergence of the open coronal magnetic field lines measured by the expansion factor, $f$, and the presence of a web of magnetic field quasi separatrix layers quantified in terms of the squashing factor, $Q$.
Both factors are derived from the 3D MHD model of the global corona.

\subsection{\label{sec:density_results}Coronal density results}

\begin{figure}
    \centering
    \includegraphics[trim=0 0 0 2cm,clip,width=\columnwidth]{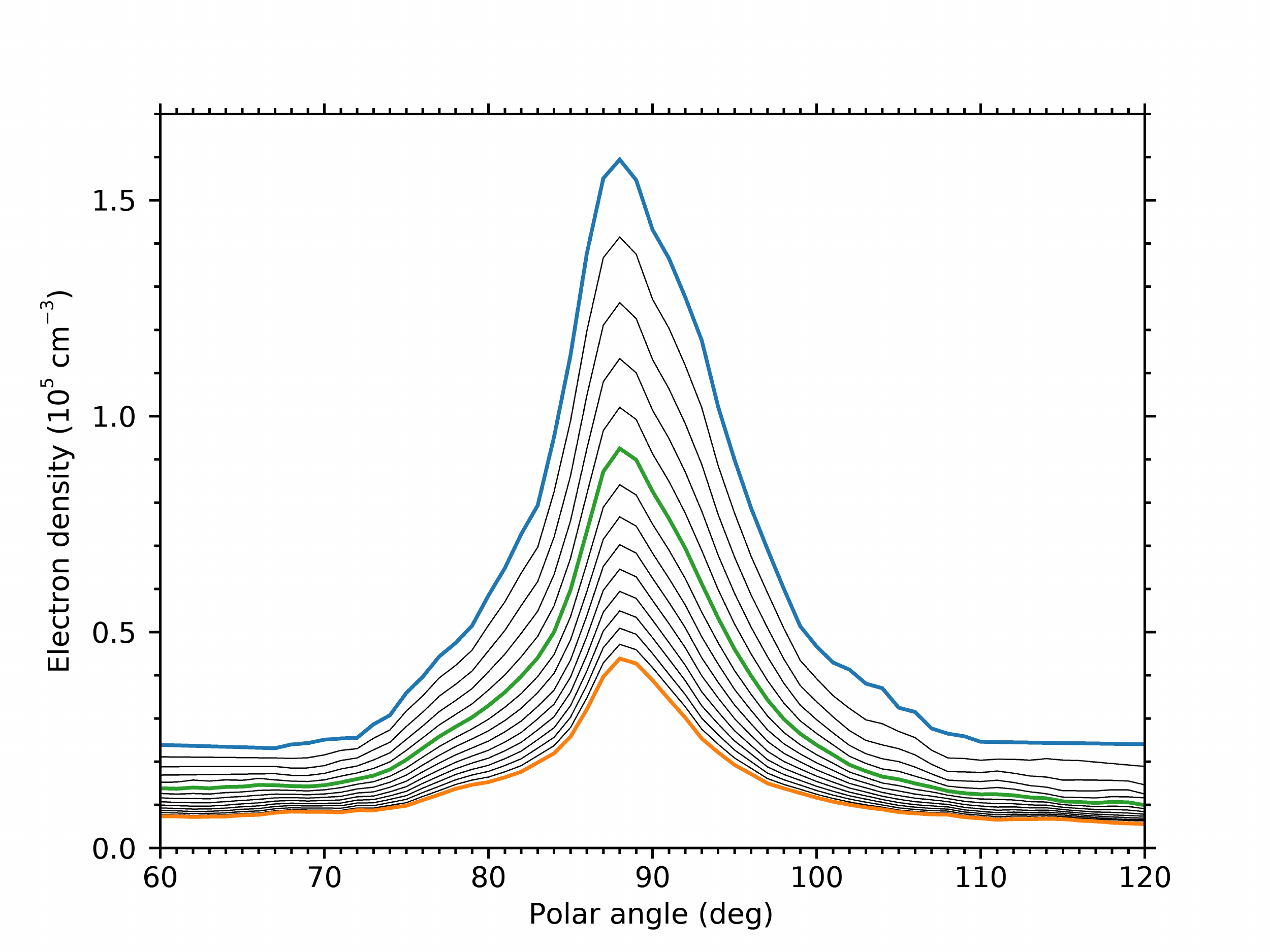}
    \includegraphics[trim=0 0 0 2cm,clip,width=\columnwidth]{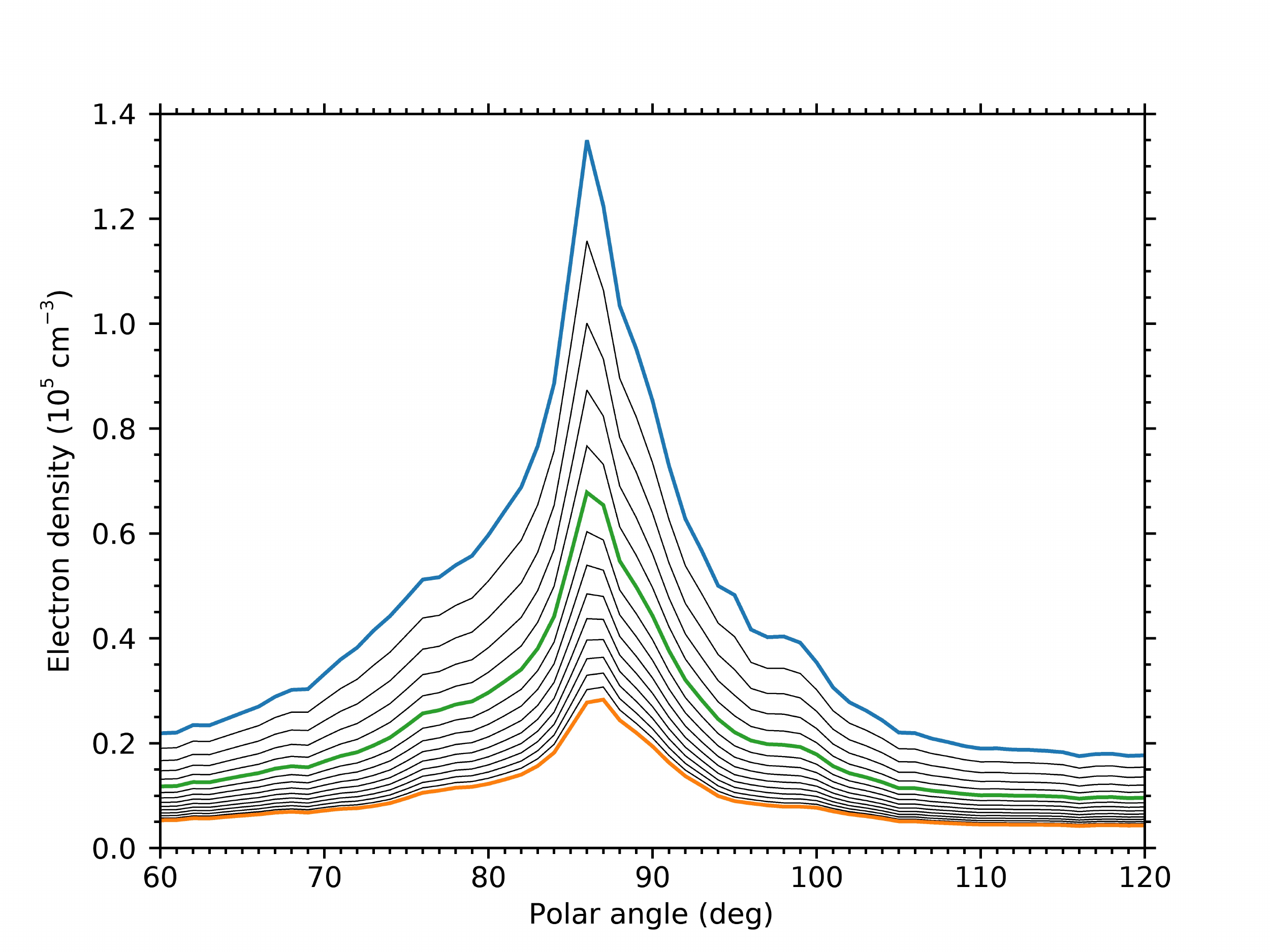}
    \caption{\label{fig:5}Latitudinal profiles of the electron density derived from the observed polarization brightness, $pB$, in the assumption of: 1) cylindrical symmetry of the coronal density (upper panel), and 2) longitudinal dependence of this quantity resulting from the 3D MHD model (lower panel). The latitudinal profiles are reported in steps of 0.2~$R_\odot$ (blue line at 4.0~$R_\odot$, green line at 5.0~$R_\odot$, orange line at 6.8~$R_\odot$).}
\end{figure}

The observed polarized brightness latitudinal profiles (Figure~\ref{fig:3}), outlining the electron density structure of the corona, show a well-defined quasi-symmetric layer of denser plasma centered on the projection on the plane of the sky of the surface (shown in Figure~\ref{fig:2}a) dividing positive and negative polarity field lines of the quasi-dipolar coronal magnetic field.
Within this layer the density decreases rapidly from the peak to the $1/e$ value within a latitudinal span of approximately $\pm10^\circ$ from the equator.
The density profiles obtained in the assumption of cylindrical symmetry (Figure~\ref{fig:5}, upper panel) outline a well-defined plasma sheet showing not much evident lateral structures.
On the axis of the sheet, few degrees North relative to the equator, the density varies from $1.6\times10^5$~cm$^{-3}$ to $0.45\times10^5$~cm$^{-3}$ at 4.0~$R_\odot$ and 6.8~$R_\odot$, respectively.
In the lateral wings extending approximately from $\pm10^\circ$ to $\pm30^\circ$ the density flattens forming quasi-plateaus, with the northern wing more structured than the southern one.
If the electron density is derived taking into account the polarized emissivity functional dependence along the LOS derived from the MHD model (Figure~\ref{fig:5}, lower panel), the density radial gradient along the axis of the plasma sheet is slightly less pronounced ($1.35\times10^5$~cm$^{-3}$ to $0.3\times10^5$~cm$^{-3}$ at 4.0~$R_\odot$ and 6.8~$R_\odot$, respectively), and the lateral structures, approximately at $10^\circ$ from the axis of the plasma sheet, are more evident.
The second case should better represent the corona viewed by an instrument integrating the signal along the LOS.
In the northern hemisphere, the negative density gradient is less steep.
In addition, the presence of an active region located at $30^\circ$ latitude East of the limb may in part contribute to this asymmetric behavior.
At the edges of the $\pm30^\circ$ latitude range the electron density reaches values approaching the coronal hole ones.
The core of the density layer close to the equator corresponds to the streamer plasma sheets which are seen at the limb almost edge on beyond 6~$R_\odot$ with LASCO C3 and exhibit a latitudinal width which can be as small as $3^\circ$ \citep{wang1998}.

\subsection{\label{sec:wind_results}Solar-wind velocity results}

\begin{figure}[ht]
    \centering
    \includegraphics[trim=0 0 0 2cm,clip,width=\columnwidth]{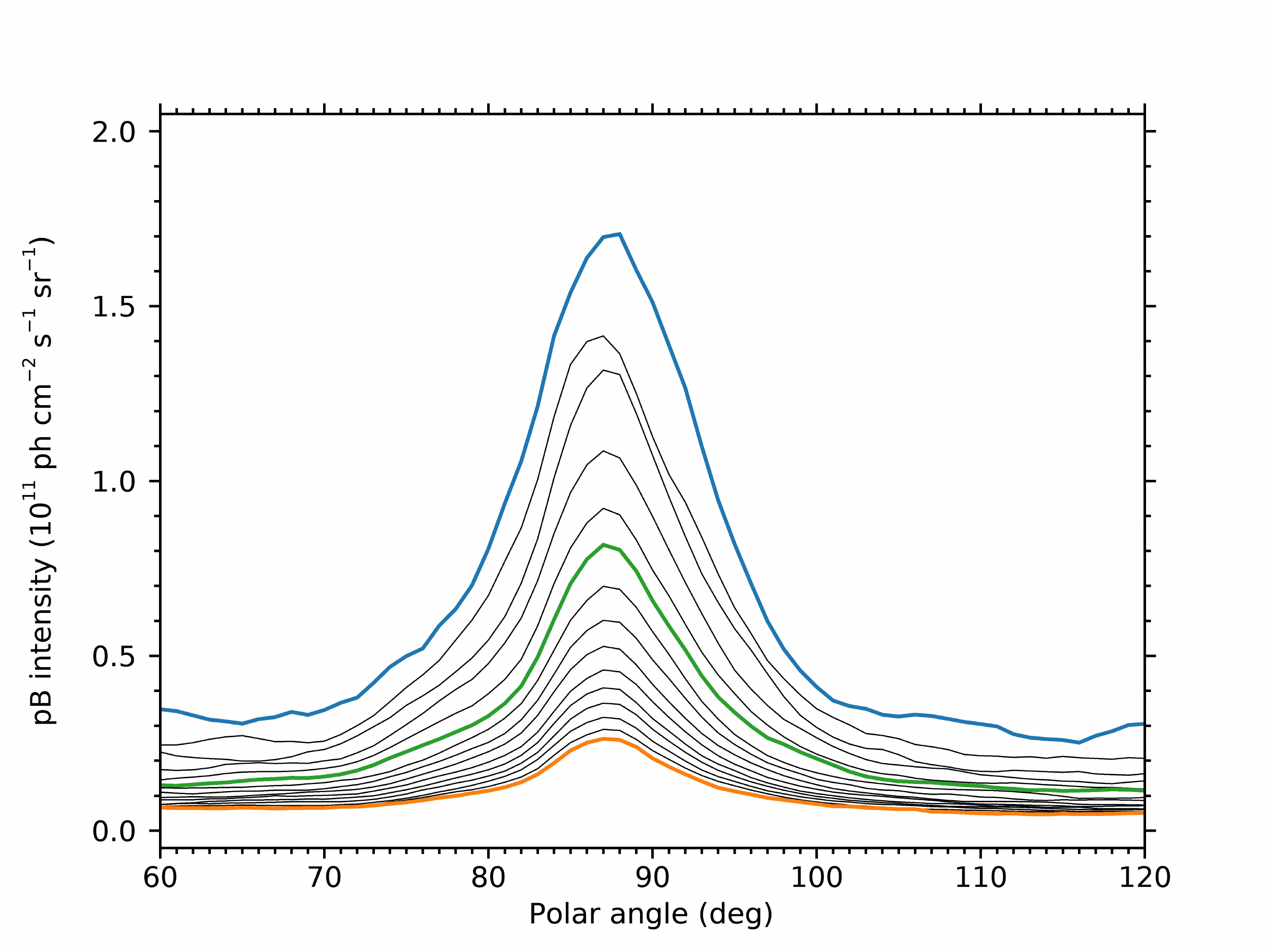}
    \includegraphics[trim=0 0 0 2cm,clip,width=\columnwidth]{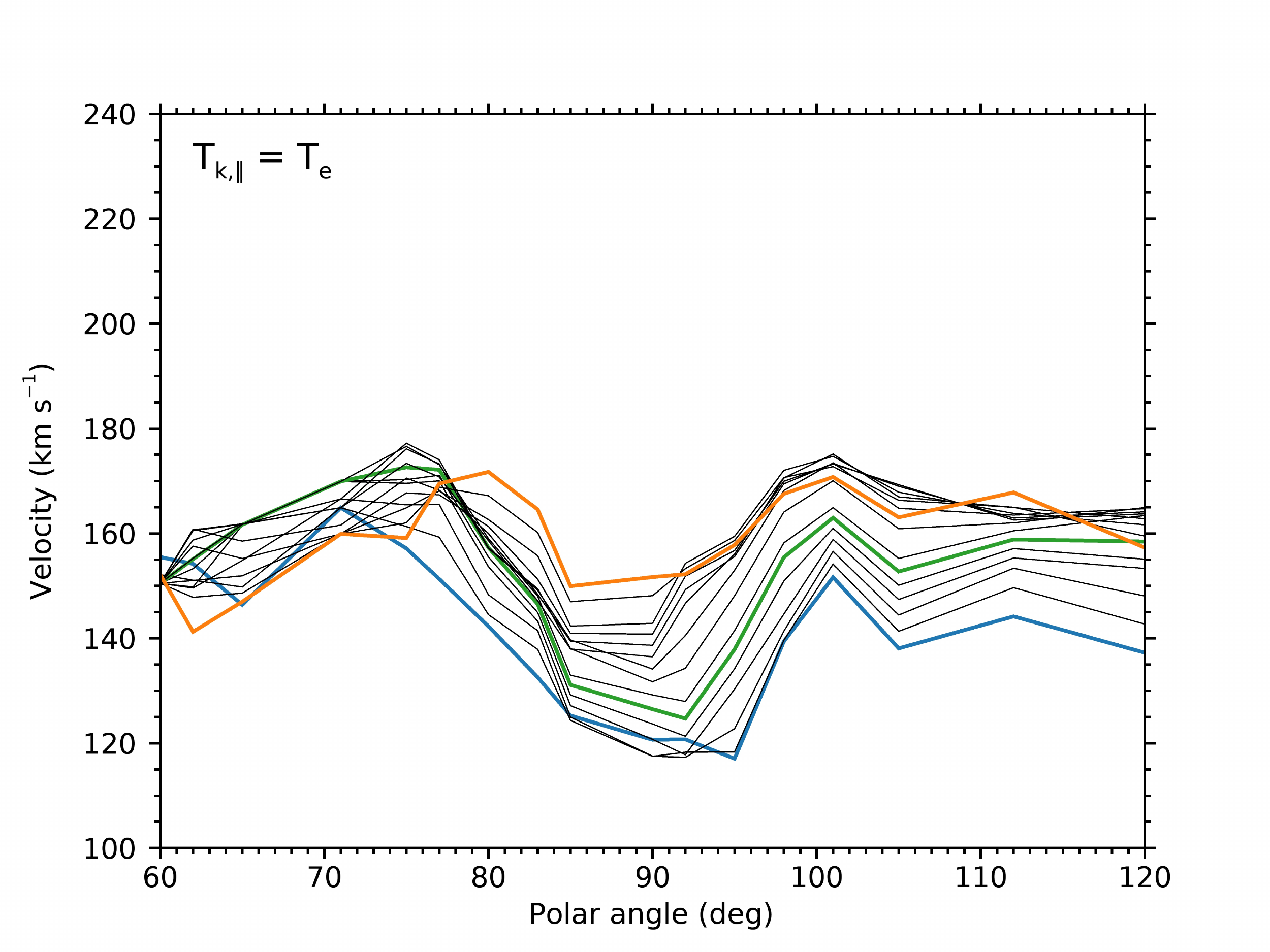}
    \includegraphics[trim=0 0 0 2cm,clip,width=\columnwidth]{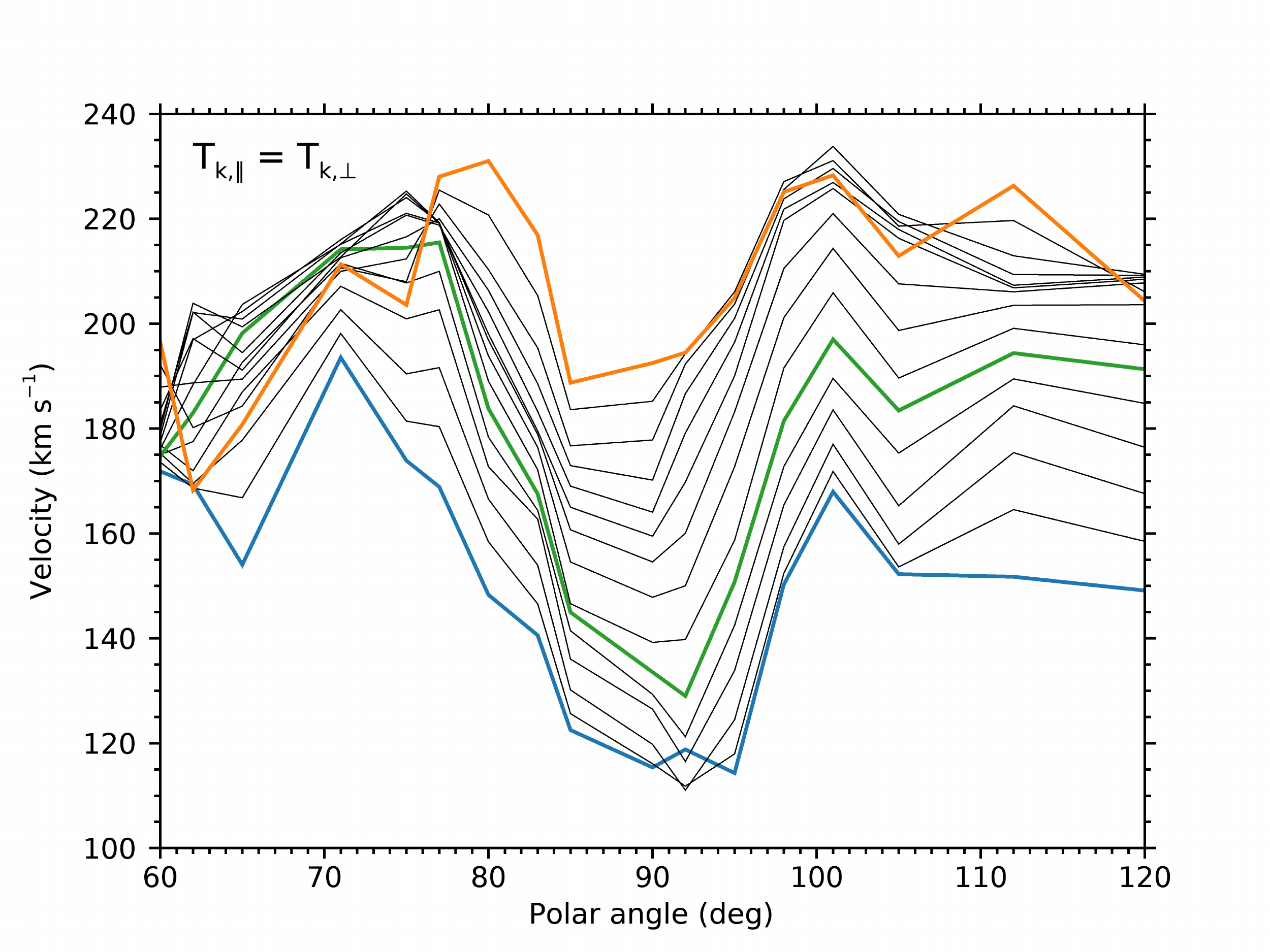}
    \caption{\label{fig:6}Upper panel: latitudinal profiles of the polarized brightness, indicated as $pB$ intensity, observed with Metis. Middle panel:  latitudinal profiles of the wind speed derived in the assumption of anisotropic distribution of the H~{\sc i} kinetic temperature ($T_{k,\perp}>T_{k,\parallel}=T_e$). Lower panel: latitudinal profiles of the wind speed derived in the assumption of isotropic distribution of the H~{\sc i} kinetic temperature ($T_{k,\perp}=T_{k,\parallel}$). The latitudinal profiles are reported in steps of 0.2~$R_\odot$ (blue line at 4.0~$R_\odot$, green line at 5.0~$R_\odot$, orange line at 6.8~$R_\odot$).}
\end{figure}

The expected anticorrelation of the speed and the electron density of the coronal outflows, according to the solar wind models and confirmed by the UVCS data, is clearly shown in Figure~\ref{fig:6}, where the observed polarized brightness (upper panel) is compared with the coronal wind velocity results obtained in the slow wind coronal zone (lower panels).
The velocity is derived in the assumption of the electron density profiles in case the model density longitudinal distribution is adopted, for anisotropic and isotropic H~{\sc i} kinetic temperature in the middle and lower panel of Figure~\ref{fig:6}, respectively.
The current sheet along the streamer axis near the equator is characterized by the slowest densest wind component.
Thin layers characterized by large velocity gradients delimit the core plasma sheet and separate the slowest wind from two, relatively faster and less dense wind flows at about $15^\circ$N and $10^\circ$S.

In the plasma sheet the wind accelerates from $\sim117$~km~s$^{-1}$ to $\sim150$~km~s$^{-1}$ and from $\sim115$~km~s$^{-1}$ to $\sim190$~km~s$^{-1}$, as it propagates from 4.0~$R_\odot$ to 6.8~$R_\odot$ above the disk surface, for anisotropic ($T_{k,\perp}=1.6$~MK, $T_{k,\parallel}=T_e$) and isotropic H~{\sc i} kinetic temperatures ($T_{k,\perp}=T_{k,\parallel}=1.6$~MK), respectively.
The velocity curves show a rapid increase from the minimum value in correspondence to the current sheet to a peak value near $15^\circ$N and $10^\circ$S, that at 6.8~$R_\odot$ is of $\approx175$~km~s$^{-1}$ and 230~km~s$^{-1}$, in the H kinetic temperature anisotropic and isotropic case, respectively.
The wind in the two streams of faster plasma reach similar peak velocity at 6.8~$R_\odot$.
In the slow stream the coronal plasma is accelerating at a significantly higher rate ($\Delta v\approx30$-75~km~s$^{-1}$ in 2.8~$R_\odot$) than in the faster wind layers surrounding the plasma sheet ($\Delta v\approx15$-40~km~s$^{-1}$ and $\Delta v\approx20$-60~km~s$^{-1}$ in 2.8~$R_\odot$, in the northern and southern higher velocity streams, respectively).
The second value of the $\Delta v$ range refers to the isotropic case.
The uncertainty on the speed due to calibration is estimated to be $\lesssim13\%$. 
The uncertainty increases toward the borders of the latitude belt under study since the signal to noise ratio decreases with distance from the equator.

\begin{figure}[ht]
    \centering
    \includegraphics[width=\columnwidth]{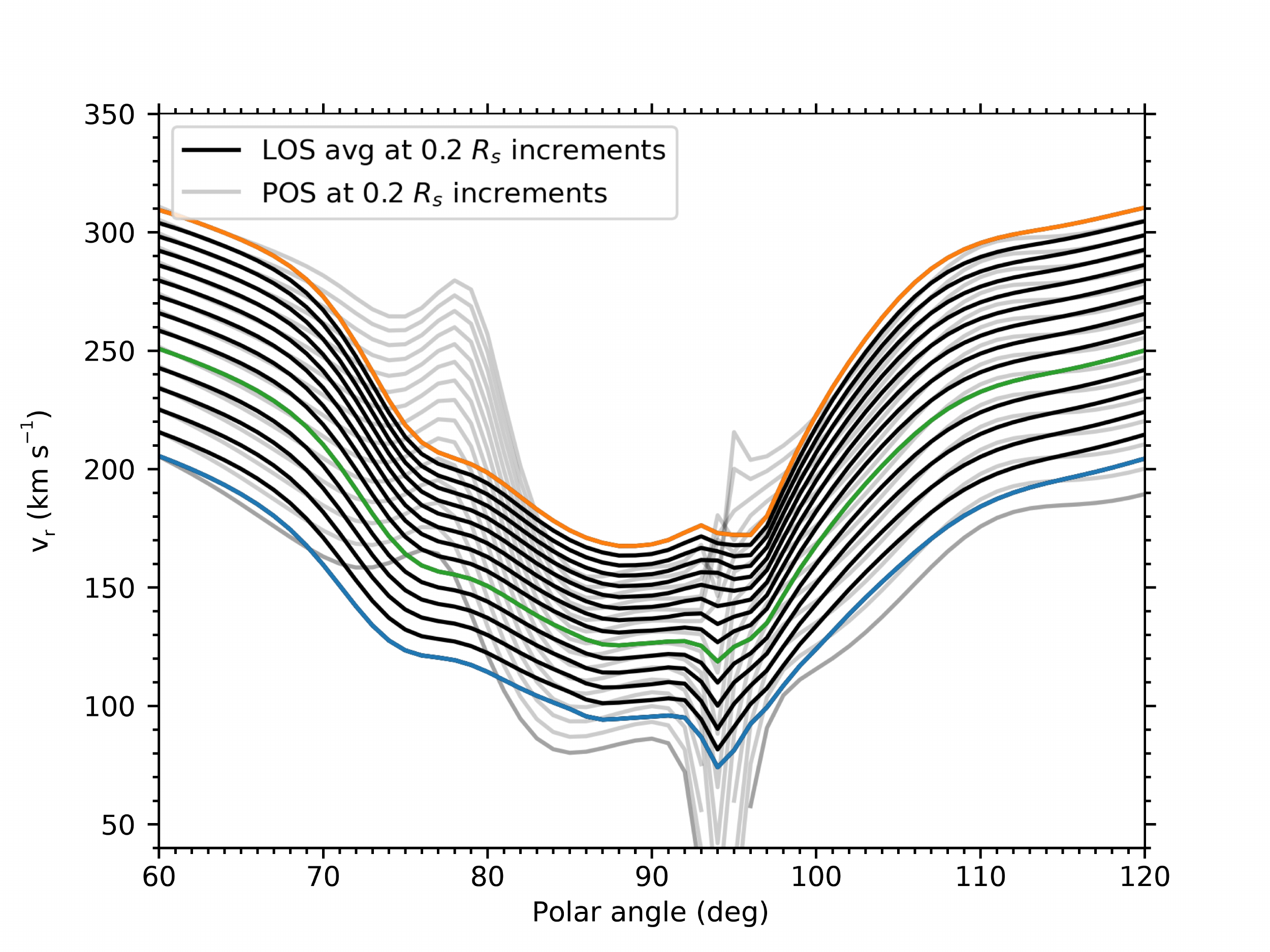}
    \caption{\label{fig:7}Latitudinal profiles of the wind velocity resulting from the MHD model: radial velocity on the plane of the sky, POS, (grey lines), and pB weighted averaged radial velocity along the line of sight, LOS, (blue line at $4.0\ R_\odot$, green line at $5\ R_\odot$, and orange line at $6.8\ R_\odot$). The latitudinal profiles are reported in steps of $0.2\ R_\odot$.}
\end{figure}

Although the discussion of the Metis data in the frame of the MHD model results has been envisaged in order to illustrate the magnetic context of the solar wind observations at coronal heights, primarily in terms of expansion factor and coronal S-web as discussed in the following sections \ref{sec:magnetic_exp} and \ref{sec:s_web}, it is also possible to derive from the model an estimate of the plane of the sky and pB weighted average radial velocity as for the other modelled physical quantities (Figure~\ref{fig:7}).
The observational and modelled wind velocities are fairly consistent within 20$^\circ$ North and 15$^\circ$ South in latitude from the equator. Whilst in the wings of the slow wind belt -- that is beyond the thin layers of sharp velocity shears surrounding the slow wind stream that is embedding the current sheet -- the wind velocity continues to increase although at a much lower rate.
This difference likely stems from the relatively simplistic way in which the solar wind is accelerated in the WTD approach, illustrating  how observations of flows in this important region for solar wind acceleration can provide useful constraints for MHD models.
The latitude of the observed fast wind streams would correspond in the case of the modelled wind to the site of a sharp change in the rate of the increase of velocity with distance from the current sheet. 
In the following section, the remarkable correlation of expansion factor and the observed wind velocity is however in favor of a slight decrease of wind velocity in the wing of the slow wind belt, as indicated by the observational results.

\section{\label{sec:magnetic_field}Magnetic-field topology in the slow-wind zone}
During sunspot minimum the slow wind is associated with the quasi-steady configuration of the corona characterized by expansion factors of the open magnetic field lines much larger in the polar coronal hole boundary regions than in its core.
Hence at least during solar minimum, the bulk of the slow solar wind is likely to originate mainly in the coronal hole peripheral regions.
This view is also supported by the fact that with the intensification of solar activity after sunspot minimum, smaller coronal holes, located at lower latitudes, become sources of slower wind.

\subsection{\label{sec:magnetic_exp}Magnetic-field super-radial expansion}
\begin{figure}[ht]
    \centering
    \includegraphics[width=\columnwidth]{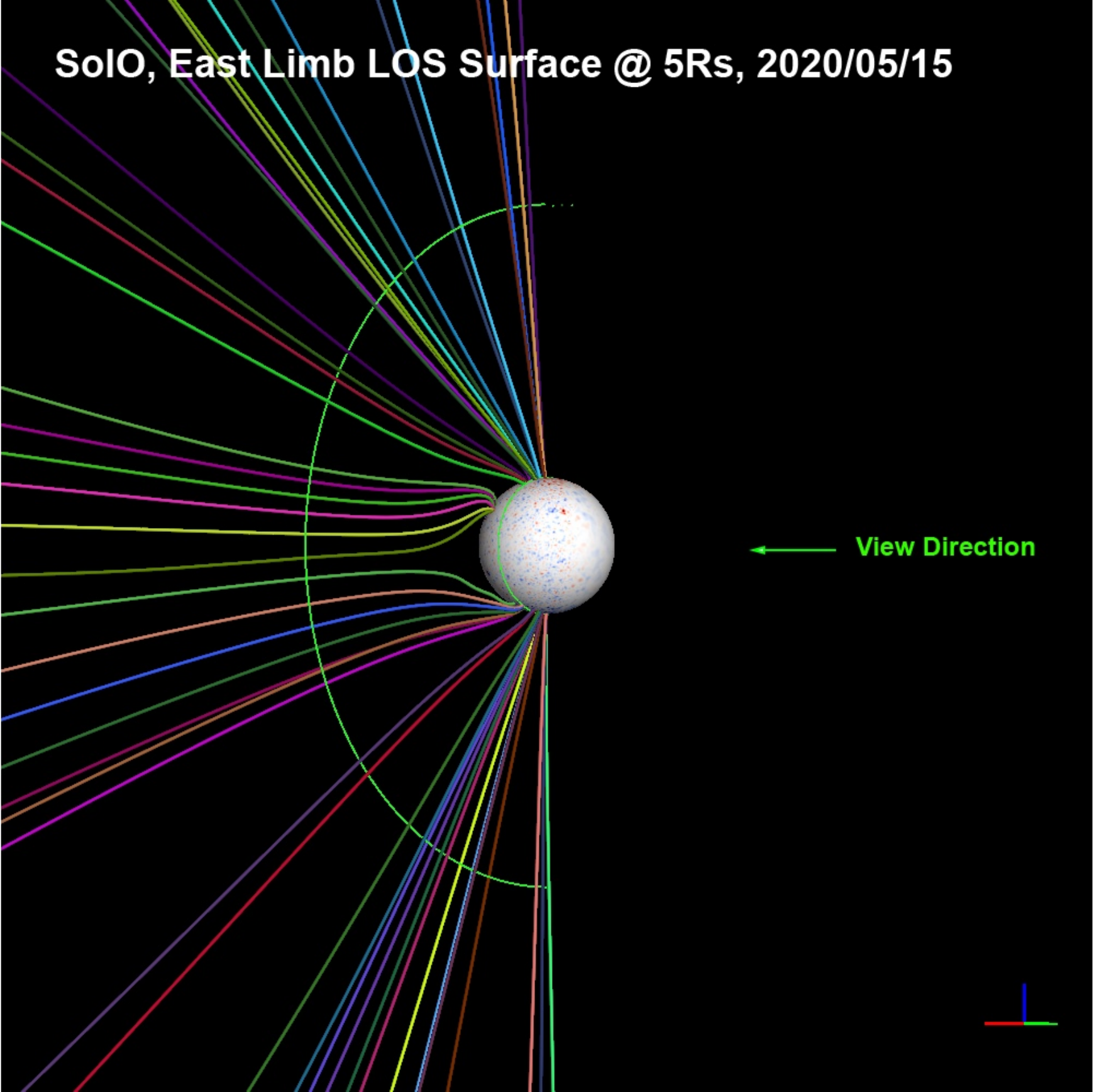}
    \caption{\label{fig:8}Modeled magnetic field lines traced from the Metis coronagraph on the plane of the sky at the East limb on May 15, 2020; the intersection with the 5~$R_\odot$ surface is shown. For perspective the view is again rotated of $45^\circ$ with respect to the Metis plane of the sky.}
\end{figure}

Let us first compare the wind speed results with the degree of super-radial expansion of the coronal magnetic field.
In the present case, due to the limited extension of the East limb streamer, the coronal hole open field lines defining the streamer boundary are converging toward the current sheet below 4~$R_\odot$, lower limit of the Metis field of view, as confirmed by the field lines computed with the MHD model that are substantially radial when these are crossing the 5~$R_\odot$ contour (Figure~\ref{fig:8}).

The expansion factor of the magnetic field is computed by taking into account the magnetic field radial component, $B_r$ \citep[e.g.,][]{wang1990,riley2015}:
\begin{equation}
    f_s = (r_0/r_1)^2 B_r(r_0,\theta_0,\phi_0)/B_r(r_1,\theta_1,\phi_1),
\end{equation}
where $(r_0,\theta_0,\phi_0)$ is the location from which the flux tube expands to another coronal location denoted by $(r_1,\theta_1,\phi_1)$.

The majority of studies that compute expansion factors do so using relatively low-resolution potential field source surface models (PFSS) and/or compute the expansion factor from the inner boundary ($r_0=1.0\ R_\odot$) to the source surface (typically $r_1=2.5$ or 2.0 ~$R_\odot$). Because the MHD model adopted in this study uses a relatively high-resolution surface boundary condition beneath the East limb, $r_0=1.02\ R_\odot$ is assumed as the inner radius to effectively decay the higher order moments of the field and smooth $B_r(r_0,\theta_0,\phi_0)$ to $\gtrsim 1^\circ$ resolution here. The heliodistance $r_1=10.0\ R_\odot$ is chosen as the maximum outer radius for the $f_s$ calculation in order to cover the heights involved in the lines-of-sight of the Metis FOV, and ensure to end well above the height where streamers open up in the MHD model. Because the heights are normalized by the $r^2$ distance ratio, any open flux tube that is nearly radial by $r=2.0$ or 2.5~$R_\odot$ will have roughly the same expansion factor at $10.0\ R_\odot$.

\begin{figure}[ht]
    \centering
    \includegraphics[trim=0 0 0 2cm,clip,width=\columnwidth]{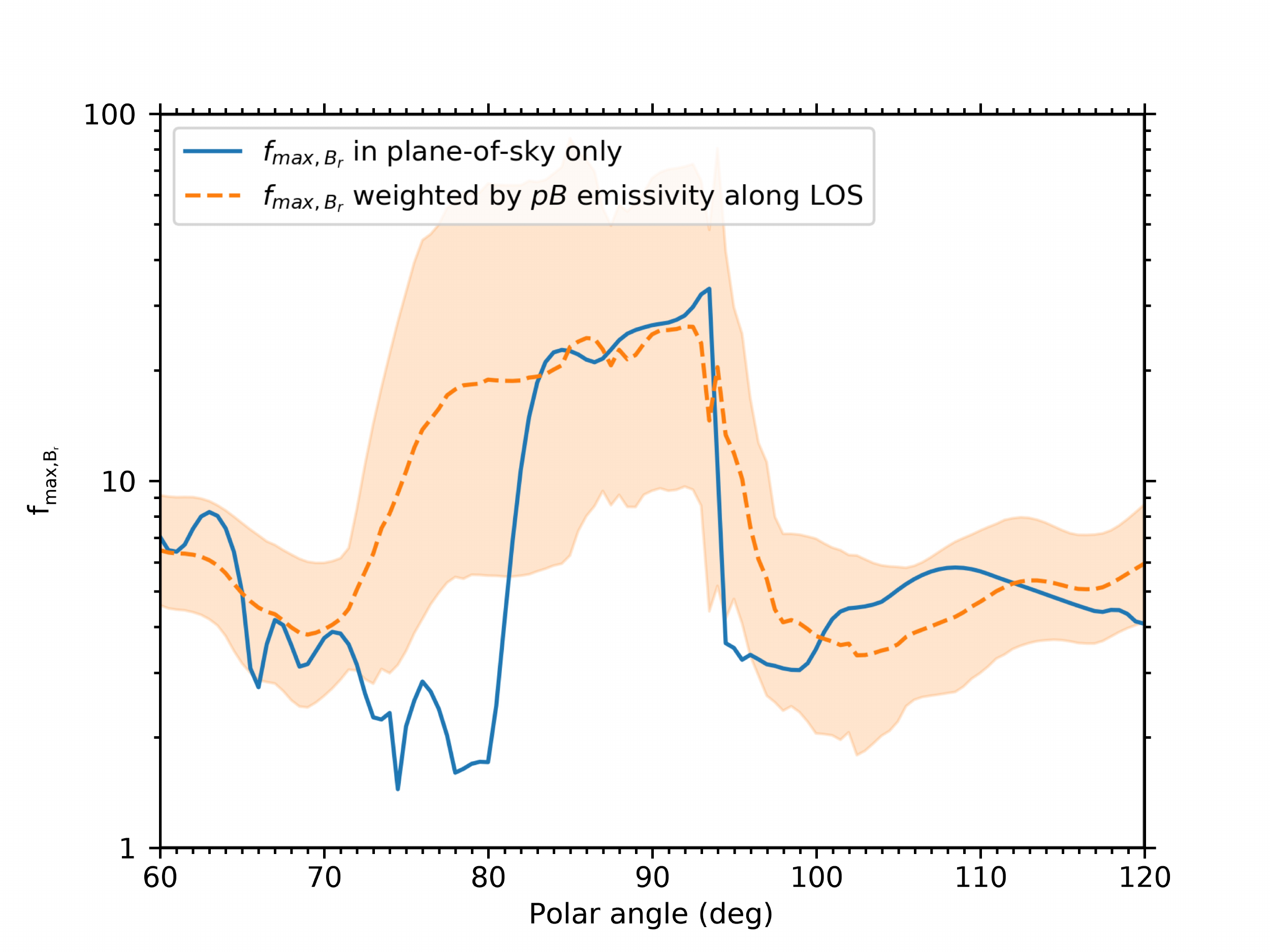}
    \includegraphics[trim=0 0 0 2cm,clip,width=\columnwidth]{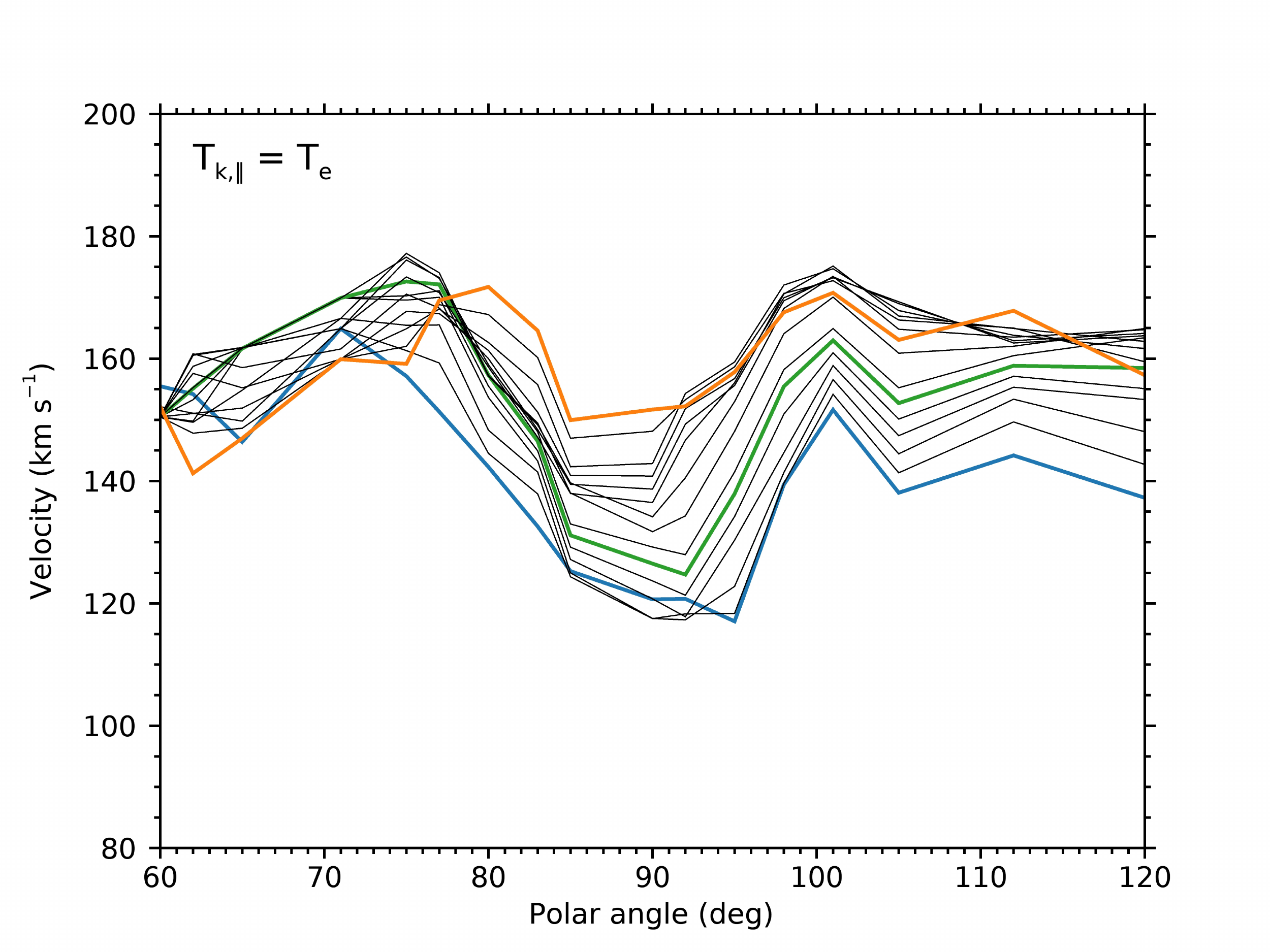}
    \includegraphics[trim=0 0 0 2cm,clip,width=\columnwidth]{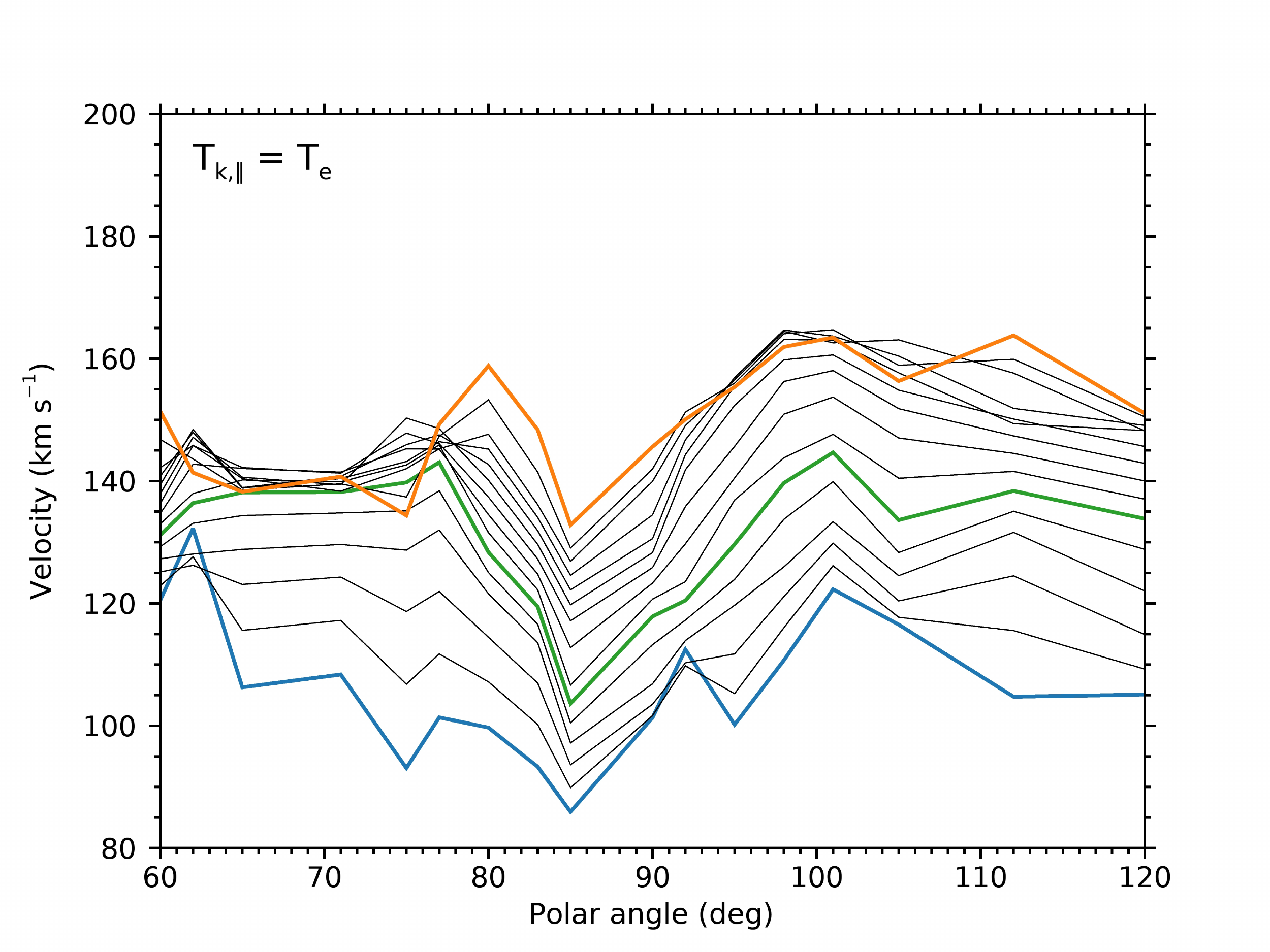}
    \caption{\label{fig:9}Upper panel: magnetic field line expansion factor, $f_{max,B_r}$, at 5~$R_\odot$, as a function of latitude at half degree intervals both for the field lines lying in the plane of the sky (continuous blue line) and for all field lines along the line-of-sight path, weighted by the polarized ($pB$) emissivity (orange dashed line). The shaded area represents the $\pm1$ standard deviation. Latitudinal profiles of the wind speed derived in the hypothesis of anisotropic distribution of the H~{\sc i} kinetic temperature ($T_{k,\perp}>T_{k,\parallel}=T_e$) for electron density derived in the assumption of a longitudinal dependence of this quantity resulting from the 3D MHD model (middle panel), and of cylindrical symmetry (lower panel), respectively. The latitudinal profiles are reported in steps of 0.2~$R_\odot$ (blue line at 4.0~$R_\odot$, green line at 5.0~$R_\odot$, orange line at 6.8~$R_\odot$).}
\end{figure}

The maximum expansion factor, $f_{max,B_r}$, experienced by the field lines between 1.02~$R_\odot$ and 10.0~$R_\odot$ is reported in the 3D plots of Figure~\ref{fig:2}b showing the surface of the LOS path at the level of 5~$R_\odot$.
This figure illustrates how the expansion factor changes rapidly on this surface.
In Figure~\ref{fig:9} (upper panel) this quantity is plotted at 5~$R_\odot$, as a function of latitude at half degree intervals both for the field lines lying in the POS (continuous line) and for all field lines along the LOS path, weighted by their contribution to the $pB$ integral (dashed line).
The mean and standard deviation are computed for the $pB$ weighted average and the shaded area behind the mean is $\pm1$ standard deviation.
The difference between the two profiles -- in the northern hemisphere the weighted expansion factor is enhanced in a more extended region -- is due to the integration of this quantity along the line of sight.

The weighted expansion factors are largest in correspondence to the polarized emissivity enhancement shown in Figure~\ref{fig:6}. The width of the latitude range exhibiting large weighted $f_{max,B_r}$ is due to  the slight warping of the almost equatorial plasma sheet observed along the LOS and peak in coincidence with the slower wind flowing at the center of the plasma sheet.
Both the maximum expansion factor and the weighted maximum expansion factor latitudinal profiles at the level of 5~$R_\odot$ (Figure~\ref{fig:9}, upper panel) show a remarkable anticorrelation with the profiles of the wind outflow speed, plotted in Figure~\ref{fig:9} (middle panel).
In addition to the feature centered at $86^\circ$, coincident with the density peak of the plasma sheet shown in Figure~\ref{fig:5}, the deepest minimum of the outflow velocity is found at $92^\circ$.
This double structure observed in the wind velocity at helio-distances $\lesssim5$~$R_\odot$ corresponds to the extended equatorial region of high expansion factor.
In the model calculations the polarized brightness and density of the plasma are necessarily related to the expansion factor; however, in the present analysis the density returned by the model has been weighted with a factor accounting for the polarized brightness observed with Metis at a given latitude.
This should ensure that the inherent correlation of flow speed and density implied by the model is not influencing the results and that the anticorrelation shown in Figure~\ref{fig:9} has a physical meaning.

The zone of large magnetic field expansion is delimited by two $f_{\mathrm{max},B_r}$ dips, at approximately $10^\circ$S and $15^\circ$N of the equator, close to the latitudes corresponding to the two fastest wind streams.
Where the expansion factor is smaller, much of the 5~$R_\odot$ connectivity is to weak quiet-sun fields that are equatorial extensions of the polar coronal holes, or, in alternative, might be related to disconnected equatorial patches.
In the wings of the latitudinal zone here considered, the expansion factors show a moderate increase which is again related to a moderate decrease in the wind velocity.
A small active region was present on the solar surface at approximately $340^\circ$ longitude and $30^\circ$N, and a flux concentration was present at the same longitude at $-25^\circ$S from the equator.
Hence the presence of the active region and the southern flux concentration are presumed to be at the origin of the $f_{\mathrm{max},B_r}$ increase, since expansion factors are larger near the small active regions and where the quiet-sun flux concentrations are strong.
On the basis of the model results, however, the computed expansion factor does not decrease toward the poles as it would be expected according to the literature which is predicting the minimum field line expansion in the core of the polar holes.
In this model, the polar caps are filled by adding many small-scale flux concentrations to represent the typical high-resolution character of surface flux, while simultaneously matching the net polar flux derived from observations \citep{mikic2018,boe2021}.
This difference might lead to an expansion factor that is slightly higher at the poles than expected from the classic papers, which use smooth polar data.

The expansion factor results are also compared with the outflow velocity resulting from the density analysis adopting the approach used in Paper~{\sc i} which does not account for the warping of the plasma sheet and uses cylindrical symmetry in the computation of the electron density derived from the observed polarized brightness.
In this case the velocity results (Figure~\ref{fig:9}, lower panel) show that the more pronounced dip in velocity is found in a narrow zone in coincidence with the axis of the plasma sheet at $86^\circ$ (Figure~\ref{fig:5}).
A second minor dip is present at least at lower heliodistances at about $95^\circ$.
In this case, the outflow velocities are better anticorrelated with the expansion in the plane of the sky only (blue line in Figure~\ref{fig:9}).

\begin{figure}[ht]
    \centering
    \includegraphics[trim=0 0 0 2cm,clip,width=\columnwidth]{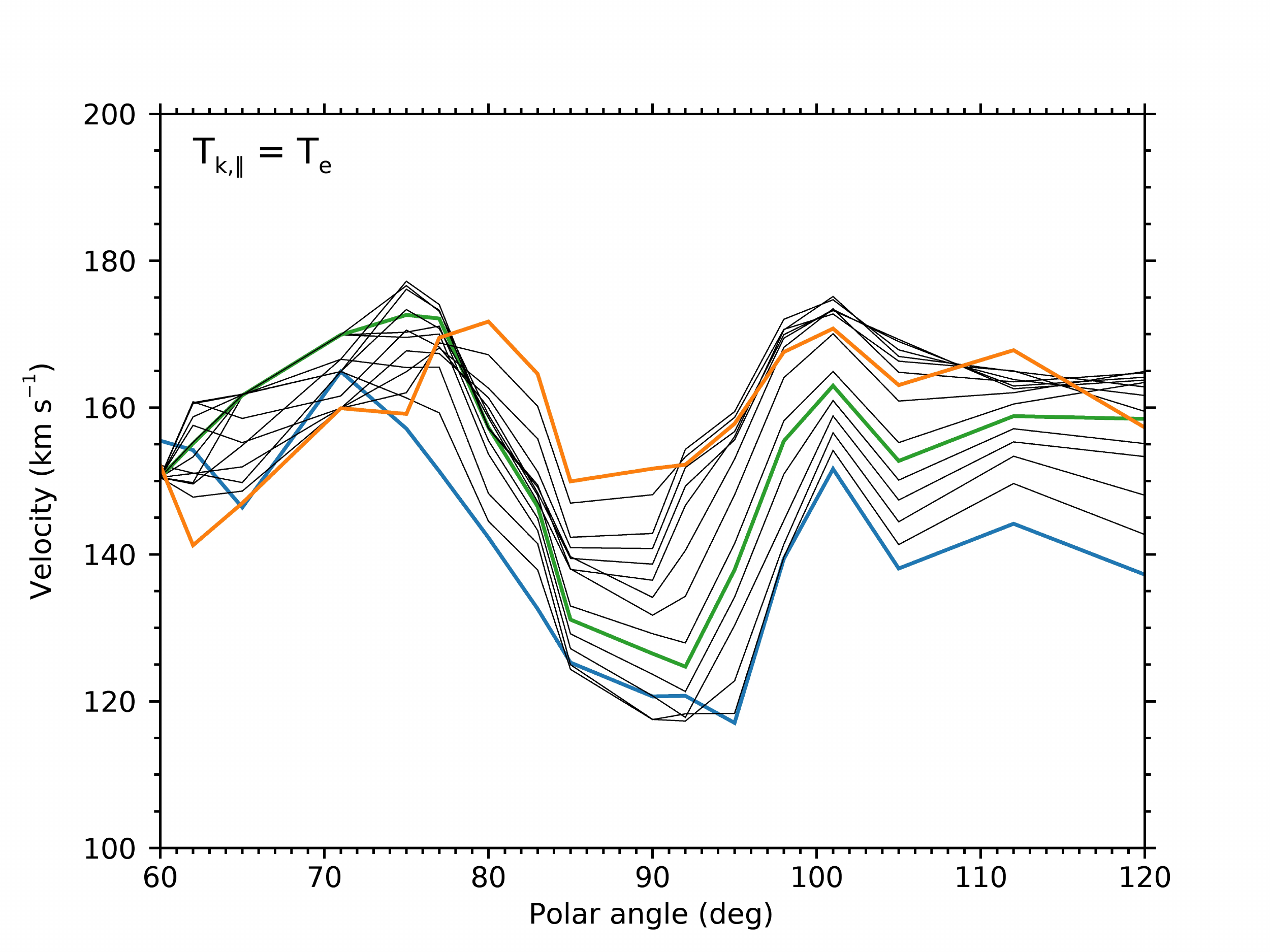}
    \includegraphics[trim=0 0 0 2cm,clip,width=\columnwidth]{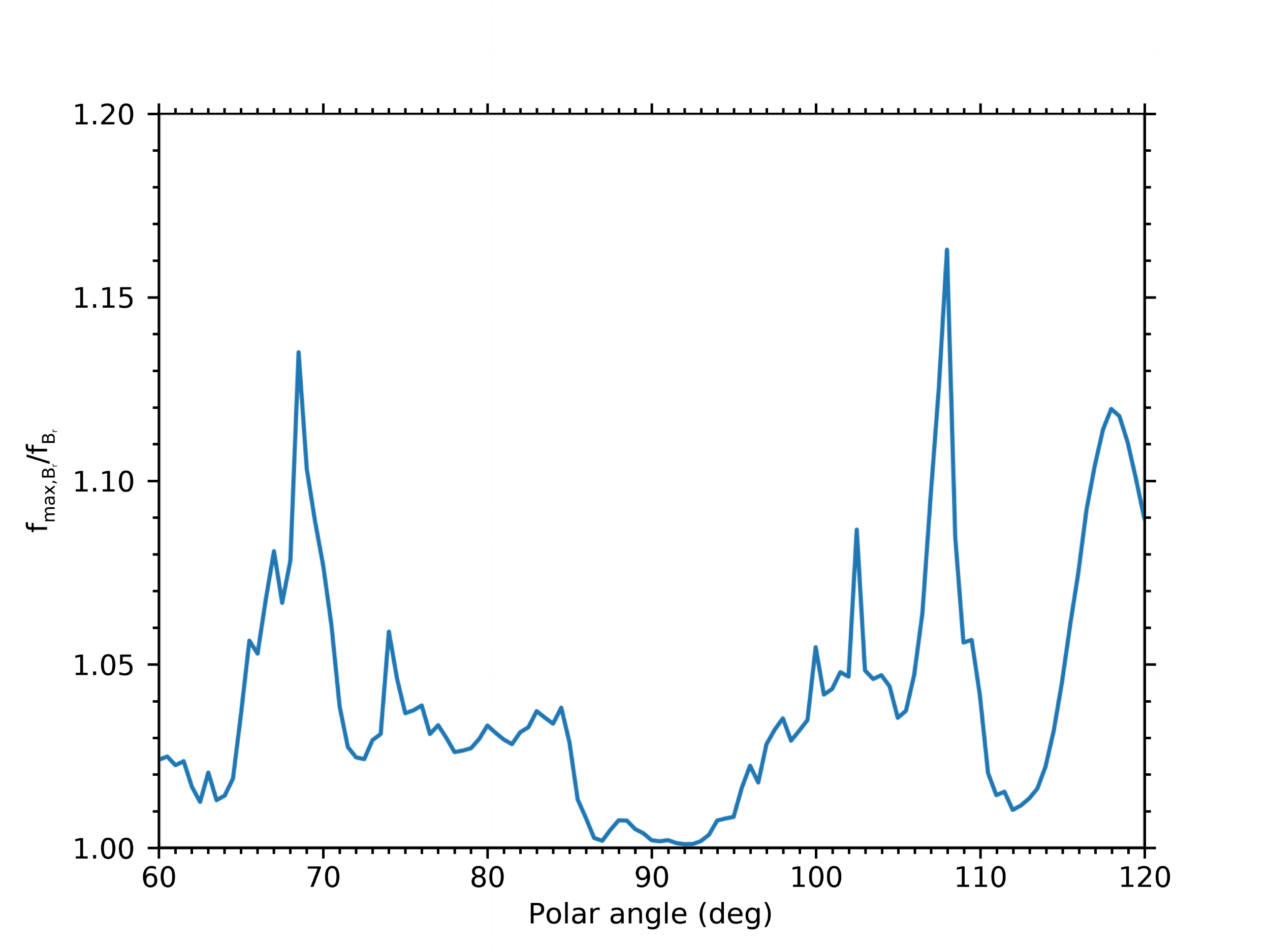}
    \includegraphics[trim=0 0 0 2cm,clip,width=\columnwidth]{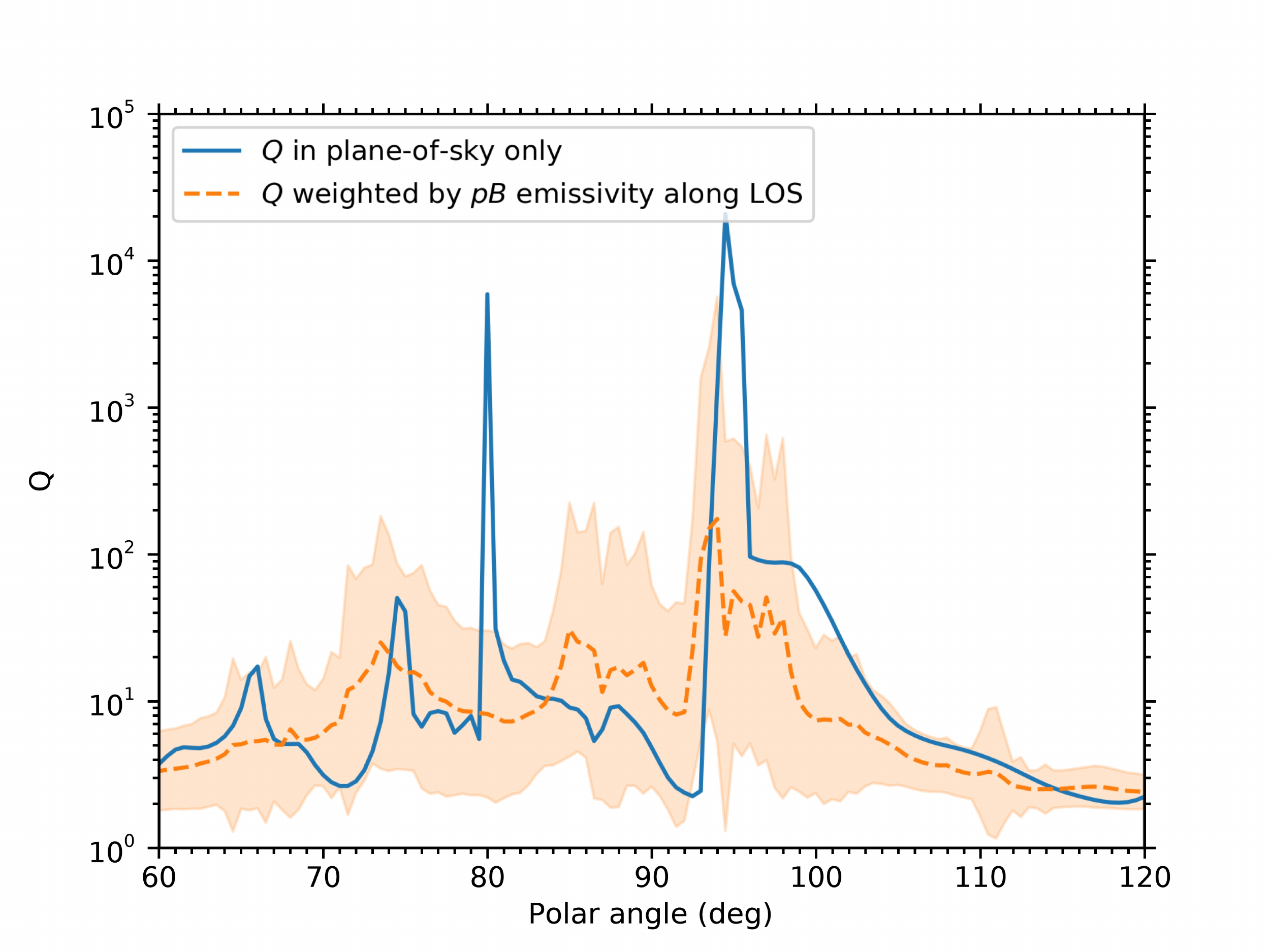}
    \caption{\label{fig:10}Latitudinal profiles of the coronal wind speed (upper panel), compared with the ratio of the maximum expansion factor of the magnetic field lines, $f_{\mathrm{max},B_r}/f_{B_r}$, relative to the expansion factor at 5~$R_\odot$ (middle panel), and the squashing factor, $Q$, in the plane of the sky (blue line) and the same factor weighted by the polarized ($pB$) emissivity along the line of sight (lower panel). The shaded area in the lower panel represents the $\pm1$ standard deviation.}
\end{figure}

The ratio of the maximum expansion factor of the magnetic field lines, $f_{\mathrm{max},B_r}/f_{B_r}$, relative to the expansion factor calculated at 5~$R_\odot$ (Figure~\ref{fig:10}) shows that the effect on the wind speed of the non-monotonic expansion ($f_{\mathrm{max},B_r}/f_{B_r}>1$) of the field lines is not so clearly delineated by the comparison of the model results and the observed wind speed.
The ratio remains always below the value 1.16 all along the East limb, indicating that the wind flux tubes are moderately non-monotonic.
At $68^\circ$ and $108^\circ$ (approximately $\pm20^\circ$ from the equator) the effect is larger, in correspondence to the negative gradient of the flow speed present in the wings of the region under study, where the speed decreases to 150~km~s$^{-1}$ (Figure~\ref{fig:10}).
This might be related to the presence of the small active region and the magnetic flux concentration at the East of the limb that is suggested to cause a moderate increase in the field expansion and a speed decrease (Figure~\ref{fig:9}).

\begin{figure}[ht]
    \centering
    \includegraphics[width=\columnwidth]{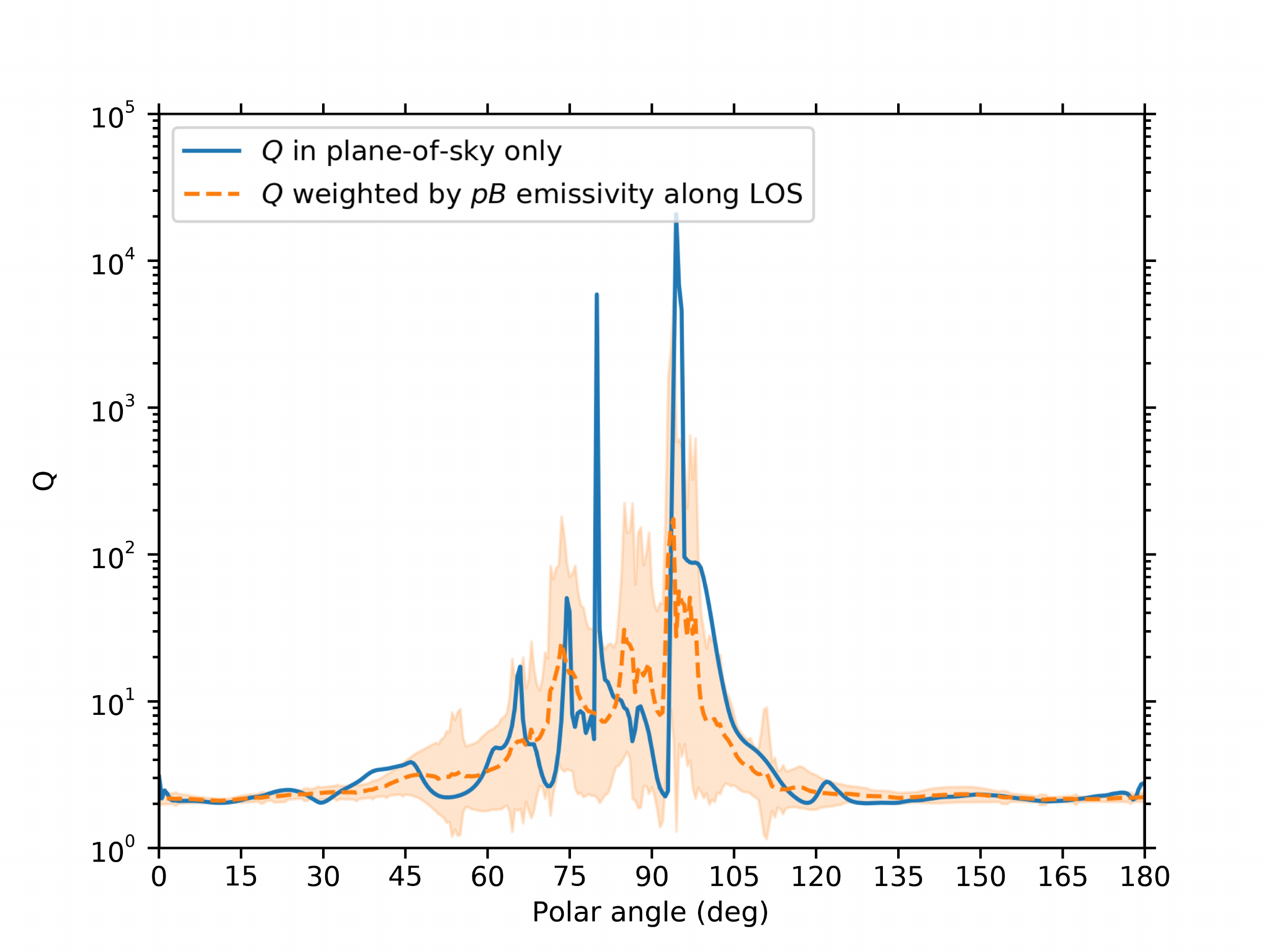}
    \caption{\label{fig:11}Squashing factor, $Q$, as a function of latitude from the North Pole ($0^\circ$) to the South Pole ($180^\circ$), in the plane of the sky (blue line) and weighted by the polarized ($pB$) emissivity along the line of sight (orange dashed line). The shaded area in the lower panel represents the $\pm1$ standard deviation.}
\end{figure}

\subsection{\label{sec:s_web}Separatrix web -- squashing factor}
The slow solar wind is also predicted to originate in the numerous open field regions or corridors in the corona, that form a web of quasi-separatrix layers, named S-Web.
The photospheric dynamics stresses these layers inducing reconnection processes with release of coronal plasma.
This hypothesis is supported by the MHD model of the corona proposed by \citet{antiochos2012}.
The separatrix web can be visualized by computing the squashing factor, Q, which measures the presence of separatrix and quasi-separatrix layers in corona, and extends up to about $30^\circ$ from the coronal-heliospheric current sheet according to the paper by \citet{titov2002}.

The plane-of-the-sky/line-of-sight surface of the quantity $\log Q$ at 5~$R_\odot$ (Figure~\ref{fig:2}c) shows the separatrix web existing during the Metis May 15 observations.
According to Figure~\ref{fig:10} (lower panel), the most significant features of the squashing factor are developing within a $\pm20^\circ$ coronal layer around the equator (see also Figure~\ref{fig:11}) and the most prominent peaks appear at $80^\circ$ and $95^\circ$ (blue line).
These are the latitudes affected by the positive gradients of the wind speed delimiting the slow stream located at the center of the plasma sheet.
The contribution of the S-web to the slow wind is clearly confined in the wind region considered in this work and in particular, at least in terms of relative position, high levels of the squashing factor can be related to the increase of the outflow speed, which is however, also coinciding with the lowest expansion factors of the magnetic field lines.

The surface illustrating the distribution of the squashing factor at 5~$R_\odot$ bears some resemblance to the surface showing the ratio of $f_{\mathrm{max},B_r}/f_{B_r}$ at the same coronal height (Figure~\ref{fig:2}c and d, respectively).
At least the main structure frame is similar in the two cases, suggesting the existence of open field lines originating in the quasi-separatrix layers which are characterized by divergent-convergent behavior.
It is interesting to note that the peaks of the ratio $f_{\mathrm{max},B_r}/f_{B_r}$, denoting the regions where the divergence-convergence behavior of the field lines is more significant, are located roughly at the boundaries of the separatrix web region (Figure~\ref{fig:10}).
That is, the S-web is limited by regions where the wind flux tubes divergence shows the largest non-monotonic behavior.

\section{\label{sec:remarks}Concluding remarks}
The observations obtained with Metis-Solar Orbiter on May 15, 2020 during the activity minimum of cycle 24 reveal the structure of the slow wind in the solar corona, shaped at that time by a magnetic dipole with axis nearly perpendicular to the heliographic equator, and allow to assess the role of the degree of expansion of the open magnetic field lines in determining the wind speed.

The solar wind is analyzed in a latitudinal belt $\pm30^\circ$ wide of the outer corona centered on the almost equatorial streamer belt.
This zone corresponds with good approximation to the region where the slow wind is expected to flow according to the knowledge arising from the out-of-ecliptic heliospheric measurements of the Ulysses instruments and from the coronal observations obtained with UVCS-SOHO at the time of the first orbit of Ulysses around the Sun at the end of the solar cycle 22.
Outside this slow wind belt, in the polar coronal holes, the neutral hydrogen/proton component of the fast wind is known to flow at a speed of 300-400~km~s$^{-1}$ at about 4~$R_\odot$ \citep[e.g.,][]{cranmer1999a}.
At the end of cycle 22 the outer solar corona, characterized by a magnetic dipole aligned with the rotation axis, was closely resembling that observed with Metis during the cycle 24 minimum, a full solar magnetic cycle later.

The detailed analysis of the Metis data complemented by the 3D MHD model calculations, show that the slow wind zone consists of a central denser and slower wind stream separated from two lateral streams of faster wind by steep velocity gradients.
In the dense core layer, $<10^\circ$ wide, running along the surface that separates opposite magnetic polarities, the wind flows at the lowest speed and undergoes an acceleration significantly higher than in the surrounding layers.
At 6.8~$R_\odot$ in the core of the streamer belt the wind speed reaches a value within 150~km~s$^{-1}$ and 190~km~s$^{-1}$, depending on the adopted kinetic temperature of the neutral hydrogen across the line of sight.
The wind speed and the areal expansion of the open magnetic field are clearly anticorrelated: the denser slower stream is found at the latitude where the expansion factor is larger.
The existence of the slowest wind along the streamer axis is consistent with the observations of the very slow wind component \citep[e.g.,][]{wang2000,antonucci2005,susino2008} ascribed to the presence of the non-monotonic divergence of the field lines in the core of the  streamer belt formed by multiple sub-streamers \citep{noci2007} as well as to the contribution of plasmoids released in magnetic reconnection processes at the streamer cusp and moving along the current sheet \citep{sheeley1997}.
The Metis data show that even in the case of a single dipole streamer and in absence of observable reconnection processes, the densest and slowest feature of the coronal wind is formed by the plasma flowing along the current sheet.

The slow wind core layer is limited by thin layers, few degrees wide in latitude, sites of a steep gradient where the velocity increases up to a peak of 175~km~s$^{-1}$-230~km~s$^{-1}$ (depending on the hydrogen kinetic temperature) at 6.8~$R_\odot$, forming two faster wind streams at about $15^\circ$N and $10^\circ$S.
Along these lateral streams the wind radial acceleration is significantly lower than in the dense and slow quasi-equatorial stream.
The faster streams are observed where the maximum expansion factor computed along the magnetic field lines, $f_{\mathrm{max},B_r}$, attains its minimum value.
In correspondence to the dips in the expansion factor latitudinal distribution, much of the 5~$R_\odot$ connectivity is to weak quiet-sun fields that are equatorial extensions of the polar coronal holes, or, in alternative, might be related to disconnected equatorial patches.
Beyond the faster wind streams, in the wings of the slow wind zone, the speed is moderately decreasing as the field line expansion is moderately increasing, with a slightly different behavior in the two hemispheres.
This effect is probably due to the presence at the surface of the Sun of a small active region at $30^\circ$N and a magnetic flux concentration at $25^\circ$S at East of the limb itself, exhibiting a larger expansion of the magnetic field lines.
Future ad hoc observations of the regions of high velocity shear between streams of solar wind flowing at different speed, as those here identified in the slow wind belt, might cast light on the energy transfer from the higher to the lower speed streams that is likely to occur and might explain the higher acceleration observed in the slow wind flowing along the current sheet.

In summary, the Metis observations show that the quasi-steady conditions of the solar wind in corona, and as a consequence the structure of the slow wind belt, are essentially regulated by the degree of the open magnetic field expansion in the corona.
Indeed, the slowest wind in the plasma sheet is spatially related to the regions of maximum super-radial divergence of the wind flux tubes.
This region is surrounded by thin streams characterized by the fastest wind observed within the zone object of the present analysis, related to the least divergent flux tubes.
The effect on the wind speed of the non-monotonic expansion of the field lines is not so clearly delineated by the comparison of the MHD model results and the observed wind speed.

According to the model results on the squashing factor, a measure of the presence of quasi-separatrix layers where reconnection processes are likely to occur in response to the photosphere dynamics with consequent plasma release, the quasi-separatrix web extends over a latitude range coincident with the $\pm30^\circ$ zone where the slow wind is observed to flow.
Furthermore, the separatrix web is delimited by magnetic flux tubes with the highest degree (maximum value of $f_{\mathrm{max},B_r}/f_{B_r}$) of non-monotonic expansion.
Hence, in the slow wind zone where the wind speed is regulated by the magnetic field areal expansion, we would also expect a significant contribution of plasma ejected during reconnection processes occurring both within the open field corridors characterizing the quasi-separatrices web \citep[e.g.,][]{antiochos2012} and in the form of plasmoids detaching close to the cusp of coronal streamers \citep[e.g.,][]{wang2000,viall2015,sanchez-diaz2017} or in the form of magnetic switchbacks \citep{telloni2022a} due to interchange reconnection.
According to this scenario, even in quiet conditions of the solar corona, when the active region contribution to the slow wind is limited or quite localized, the characteristic temporal and spatial variability of the slow wind, and its ionic and elemental composition, are ensured by the plasma ejected in the background quasi-steady slow wind during the reconnection processes.
It is interesting to note that in this case the plasma, due to magnetic reconnection events, is injected into the quasi-steady slow wind approximately in the region where the areal expansion of the magnetic field is such to give origin to the slow wind.
The assessment of the relative importance of the contributions by part of the different processes involved in the genesis of the slow wind is deferred to future ad hoc studies on the basis of combined remote sensing observations of the corona and in situ measurements of the coronal/heliospheric plasma obtained with instruments exploring the wind plasma close to Sun.

\begin{acknowledgments}
    The Metis coronagraph of the Solar Orbiter mission is primarily devoted to the study of the Parker's solar wind in the corona.
    Solar Orbiter is a space mission of international collaboration between ESA and NASA, operated by ESA.
    Metis has been built and is operated with funding from the Italian Space Agency (ASI), under contracts to the National Institute of Astrophysics (INAF) and industrial partners.
    Metis has been built with hardware contributions from Germany (Bundesministerium für Wirtschaft und Energie through DLR), from the Czech Republic (PRODEX) and from ESA.
\end{acknowledgments}

\bibliography{biblio}
\end{document}